%% file: EXOT-2012-25.tex
\documentclass[11pt,a4paper]{article}
\usepackage{jheppub}
\usepackage{atlasphysics}
\usepackage{graphicx}
\usepackage{subfigure}
\usepackage{mathrsfs}
\usepackage{preprintcover} 
\usepackage{url}
\usepackage{authblk}
\usepackage{ssdefs}
\usepackage{multirow}
\usepackage{longtable}
\usepackage{import}
\usepackage{caption} 
\usepackage{array} 
\usepackage{float}
\usepackage{afterpage}

\PreprintCoverPaperTitle{\boldmath Search for anomalous production of prompt same-sign lepton pairs and pair-produced doubly charged Higgs bosons with $\sqrt{s} = 8$~\tev\ $pp$ collisions using the ATLAS detector}
\PreprintIdNumber{CERN-PH-EP-2014-158}
\PreprintCoverAbstract{
A low-background inclusive search for new physics in events with same-sign dileptons is presented. The search uses proton--proton collisions 
corresponding to 20.3~\ifb\ of integrated luminosity taken in 2012 at a centre-of-mass energy of 8 \tev\ with the ATLAS detector at the LHC. 
Pairs of isolated leptons with the same electric charge and large transverse momenta of the type \ee, \emu, and \mumu\ are selected and their
 invariant mass distribution is examined. No excess of events above the expected level of Standard Model background is found. 
The results are used to set upper limits on the cross-sections for processes beyond the Standard Model. Limits are placed as a function of th
e dilepton invariant mass 
within a fiducial region corresponding to the signal event selection criteria. Exclusion limits are also
derived for a specific model of doubly charged Higgs boson production.
}
\PreprintJournalName{JHEP}

\title{\boldmath Search for anomalous production of prompt same-sign lepton pairs and pair-produced doubly charged Higgs bosons with $\sqrt{s} = 8$~\tev\ $pp$ collisions using the ATLAS detector}
\collaboration{The ATLAS Collaboration}

\abstract{
A low-background inclusive search for new physics in events with same-sign dileptons is presented. The search uses proton--proton collisions corresponding to 20.3~\ifb\ of integrated luminosity taken in 2012 at a centre-of-mass energy of 8 \tev\ with the ATLAS detector at the LHC. 
Pairs of isolated leptons with the same electric charge and large transverse momenta of the type $e^{\pm}e^{\pm}, e^{\pm}\mu^{\pm}$, and $\mu^{\pm}\mu^{\pm}$ are selected and their invariant mass distribution is examined. No excess of events above the expected level of Standard Model background is found. 
The results are used to set upper limits on the cross-sections for processes beyond the Standard Model. Limits are placed as a function of the dilepton invariant mass 
within a fiducial region corresponding to the signal event selection criteria. Exclusion limits are also
derived for a specific model of doubly charged Higgs boson production.
}

\begin{document}
\maketitle
\flushbottom

\section{Introduction}
\input{Introduction}

\section{The ATLAS detector}
\label{sec:det}
\input{Detector}

\section{Background and signal simulation}
\label{sec:datasets}
\input{Datasets}

\section{Physics object reconstruction}
\label{sec:particles}
\input{Particles}

\section{Data and event selection}
\label{sec:selection}
\input{Selection}

\section{Background estimation}
\label{sec:background}
\input{Background}

\section{Systematic uncertainties}
\label{sec:uncertainties}
\input{Uncertainties}

\section{Results and interpretation}
\label{sec:results}
\input{Results}

\section{Conclusion}
\label{sec:conclusions}
\input{Conclusions}

%
\acknowledgments
\input{Acknowledgement}



\bibliographystyle{JHEP}
\bibliography{EXOT-2012-25}{}
\newpage
\input{atlas_authlist.tex}

\end{document}

%% file: Introduction.tex
In proton--proton ($pp$) collisions, Standard Model (SM) processes rarely produce two isolated leptons with large transverse momentum (\pt) and the same electric charge (same-sign).
However, such signatures frequently occur in models of physics beyond the SM.
Supersymmetry~\cite{theorySUSY}, universal extra dimensions~\cite{theoryUED},  left-right symmetric models~\cite{LRSM1,LRSM2,LRSM3,LRSM4}, seesaw models~\cite{Muhlleitner:2003me,Akeroyd:2005gt,Hektor:2007uu,Perez:2008zc,Chao:2008mq,foot,Franceschini:2008pz,delAguila:2008cj}, vector-like quarks~\cite{Chivukula:1999prd,Anastasiou:2009prd,Kong:2012jhep,Babu:1990prd,Grinstein:2010jhep,Carena:2007prd}, the Zee--Babu neutrino mass model~\cite{ZBM1,ZBM2,ZBM3}, and the coloured Zee--Babu model~\cite{cZBM}
could all give rise to final states with two same-sign leptons. 

An inclusive search in events 
with pairs of isolated same-sign leptons is presented in this paper. The dilepton pairs are selected in the $pp$ collision data corresponding to 20.3~\ifb\ of integrated luminosity taken in 2012 at a centre-of-mass energy of $\sqrt{s} = 8$~\tev\ with the ATLAS detector \cite{detpaper} at the Large Hadron Collider (LHC). 
The same-sign dilepton pairs can be either two electrons (\ee), two muons (\mumu), or one electron and one muon (\emu), and must have a transverse momentum (\pt) of at least 25 (20) \gev\ for the leading (subleading) lepton. After selection of these pairs
the resulting
invariant mass distributions are examined. 
The data are found to be consistent with the SM background predictions, and exclusion limits are set on the fiducial cross-section of new physics in the same-sign dilepton final state. 
Limits are also provided separately for two positively or negatively charged leptons as a function of the dilepton invariant mass. The analysis, using the 8 TeV dataset, provides
significantly stronger constraints on new physics models
than that presented in earlier ATLAS publications using 4.7 \ifb\ of $pp$ collision data recorded at $\sqrt{s} = 7$~TeV~\cite{2011SS,2011dch}. 
Exclusion limits are also presented for the mass of pair-produced doubly charged Higgs bosons (\hpp) \cite{2011dch}.
The CDF experiment has performed similar inclusive searches \cite{cdf1, cdf2} without observing any evidence for new physics.  
This search is more inclusive than other similar searches 
at ATLAS and CMS in events with same-sign dileptons with additional requirements on missing transverse energy, jets, and charged particles \cite{cmsHeavyB, cmsSUSY, cmsMajorana, atlasSusySS, atlasSStop, atlasStrongGravity, cmsdch}. 
 Recently, the ATLAS experiment published limits on doubly charged Higgs production in multi-lepton events~\cite{atlasMultilep} based on the $\sqrt{s} = 8$~\tev\ data.

%% file: Detector.tex
From the inside to the outside, the ATLAS detector comprises an inner 
tracking detector (ID), electromagnetic and hadronic calorimeters, and a muon spectrometer (MS). The ID is embedded in a 2 T axial magnetic field produced by a superconducting solenoid and 
provides precision tracking within the pseudorapidity\footnote{ATLAS uses a right-handed coordinate system with its origin at the nominal interaction point in the centre of the detector and the $z$-axis along the beam line. The $x$-axis points from the interaction point to the centre of the LHC ring, and the $y$-axis points upwards. Cylindrical coordinates ($r$, $\phi$) are used in the transverse plane, $\phi$ being the azimuthal angle around the beam line. The pseudorapidity is defined in terms of the polar angle $\theta$ as $\eta = - \ln \tan(\theta/2)$.} range $|\eta|<2.5$. It consists of a silicon pixel detector, a semiconductor tracker (SCT) using silicon microstrip detectors, and, in the region $|\eta|<2$, a transition-radiation straw tube tracker (TRT).

The calorimeter system consists of electromagnetic and hadronic components and covers the pseudorapidity range $|\eta|<4.9$.
The electromagnetic calorimeter is a lead/liquid-argon sampling calorimeter. 
It covers $|\eta|<3.2$ with a fine lateral and longitudinal segmentation up to $|\eta|=2.5$, and is subdivided into a barrel ($|\eta|<1.4$) and two endcaps ($1.5<|\eta|<3.2$). 
The steel/scintillator--tile hadronic calorimeter provides coverage up to $|\eta|=1.7$, while the hadronic calorimeter in the endcap ($1.5<|\eta|<3.2$) and in the forward region ($3.1<|\eta|<4.9$) uses liquid argon technology.

The muon spectrometer uses toroidal magnetic fields generated by three large superconducting magnet systems with eight coils each. The detector is made up of separate trigger and high-precision tracking chambers. The precision chambers cover the region 
$|\eta|<2.7$ with three layers of monitored drift tube chambers, complemented by cathode strip chambers in the forward region. The trigger system covers the range $|\eta|<2.4$ using resistive plate chambers in the barrel and thin-gap chambers in the endcap regions. 

A three-level trigger system is used to select events. The first level is implemented in custom electronics, and is followed by two software-based trigger levels. This system selects from the collision rate of around 20 MHz about 400 Hz of events to be recorded for physics analyses. 

More details about the detector and the trigger system can be found elsewhere \cite{detpaper}.

%% file: Datasets.tex
Monte Carlo (MC) simulations are used to estimate the background contributions and also to model hypothetical signal events. 
The MC background samples used are shown in table \ref{tab:MC_cross}. For each process the table provides information on the generator, the chosen parton distribution function (PDF) and the order of cross-section calculations used for the normalisation. 

The irreducible background in the analysis comes mainly from the purely leptonic decays of $WZ$ and $ZZ$ production processes. A small contribution arises
from $\Wpm\Wpm$ production \cite{wwjj}, which proceeds 
via the t-channel exchange of a gluon and results in at least two jets in the final state, in addition to the two $W$ bosons.
The small contributions from multiple parton interactions (MPI) for $WW$, $WZ$ and $ZZ$ are also considered. 
In these processes two hard scatterings occur in the same $pp$ collision, each producing either a $W$ or $Z$ boson. 
Other smaller sources of background are the processes in which a $W$ or $Z$ boson is produced in association with a top-quark pair (\ttbar $W$ and \ttbar $Z$).
One of the reducible backgrounds arises from the opposite-sign lepton pairs where the charge for one of the leptons is wrongly reconstructed. 
In order to estimate the contribution, Drell--Yan ($Z/\gamma^*+$jets), \ttbar, \Wpm \Wmp\, and $Wt$ simulations are used, and 
misidentification probabilities derived from data are applied to the MC samples (see section~\ref{sec:oppsignbkg}).
The process $W\gamma$, where the photon converts to an $e^+ e^-$ pair, is also simulated. The production of $Z\gamma$ is included in the $Z / \gamma^*$ process. 

The MC program {\sc sherpa-1.4.1} \cite{Sherpa} is used to model the $WZ$, $ZZ$, \Wpm\Wmp\ and $W\gamma$ processes.
These samples use the default {\sc sherpa} parameterisation for the renormalisation and factorisation scales. 
For processes with a $Z$ boson, the contribution from $\gamma^{*} \rightarrow \ell^{+}\ell^{-}$ due to internal or external bremsstrahlung of final-state quarks or leptons is simulated for $m(\ell^{+}\ell^{-}) > 0.1$~\gev. 
The \ttbar$W$, \ttbar$Z$ and \Wpm\Wpm\ events
are generated using {\sc MadGraph-5.1.4.8} \cite{madgraph4, madgraph5},
and for the fragmentation and hadronisation, {\sc pythia-6.426} \cite{Pythia} is used for \ttbar$W$ and \ttbar$Z$ and {\sc pythia-8.165} \cite{pythia8} for \Wpm\Wpm. 
The MPI samples are generated by {\sc pythia-8.165}.
The Drell--Yan process is modelled using {\sc alpgen-2.14} \cite{Alpgen}, and the top-quark pair production and single top-quark production in association with a $W$ boson are generated with {\sc mc@nlo-4.06} \cite{Mcnlo, Mcnlo2}. These are interfaced to {\sc herwig-6.520} \cite{Herwig1, Herwig2} for the fragmentation and hadronisation process, and {\sc jimmy-4.31} \cite{Jimmy} is used for the underlying-event description.  

\begin{table}[ht]
  \begin{center}
    \begin{tabular}{l|c|c|c}
      
      \hline
Process &  Generator&  PDF set & Normalisation \\
&  + fragmentation/ &  & based on \\
&  hadronisation & &\\
\hline\hline
\multirow{2}{*}{$WZ$ } &  \multirow{2}{*}{{\sc sherpa-1.4.1} \cite{Sherpa}} &   \multirow{2}{*}{CT10 \cite{ct10}} & NLO QCD \\
 & & &  with {\sc mcfm-6.2}\cite{mcfm} \\
\hline
\multirow{2}{*}{$ZZ$}  &  \multirow{2}{*}{{\sc sherpa-1.4.1}} & \multirow{2}{*}{CT10} & NLO QCD  \\
& & &  with {\sc mcfm}-6.2 \\
\hline
\multirow{2}{*}{\Wpm\Wpm}  & M{\sc ad}G{\sc raph}-5.1.4.8 \cite{madgraph4}  &   \multirow{2}{*}{CTEQ6L1 \cite{cteq}}  &  \multirow{2}{*}{LO QCD} \\
&  {\sc pythia-8.165} \cite{pythia8}& &\\
\hline
\ttbar $V$, & M{\sc ad}G{\sc raph}-5.1.4.8  & \multirow{2}{*}{CTEQ6L1} & \multirow{2}{*}{NLO QCD \cite{top9,ttbarW}} \\
$V=W,Z$ &  + {\sc pythia-6.426} & & \\
\hline
 MPI $VV$ &  \multirow{2}{*}{{\sc pythia-8.165}\cite{pythia8}}  &  \multirow{2}{*}{CTEQ6L1} &  \multirow{2}{*}{LO QCD} \\
 $V=W,Z$ &  & & \\[+0.025in]
\hline
\hline
\multirow{2}{*}{$Z/\gamma^* +$ jets} & {\sc alpgen-2.14} \cite{Alpgen}&\multirow{2}{*}{CTEQ6L1}& {\sc dynnlo-1.1} \cite{dynnlo} with \\
 & + {\sc herwig-6.520} \cite{Herwig1, Herwig2}& & MSTW2008 NNLO \cite{mstw} \\
\hline
\multirow{2}{*}{\ttbar} & {\sc mc@nlo}-4.06 \cite{Mcnlo, Mcnlo2} & \multirow{2}{*}{CT10}&{NNLO+NNLL } \\
& + {\sc herwig-6.520} & & QCD \cite{top1,top2,top3,top4,top5,top6} \\
\hline
\multirow{2}{*}{$Wt$} & {\sc mc@nlo-4.06}  & \multirow{2}{*}{CT10}& {NNLO+NNLL } \\
   & + {\sc herwig-6.520} & & QCD \cite{top7,top8}\\
\hline
\multirow{2}{*}{$W^{\pm}W^{\mp}$} & \multirow{2}{*}{{\sc sherpa-1.4.1}} & \multirow{2}{*}{CT10}& NLO QCD \\
& &  & with {\sc mcfm-6.2}\\
\hline
\multirow{2}{*}{$W\gamma$} & \multirow{2}{*}{{\sc sherpa}-1.4.1} & \multirow{2}{*}{CT10}& NLO QCD\\
& &  & with {\sc mcfm-6.3}\\
\hline
\end{tabular}
\end{center}
  \caption{Generated samples used for background estimates. The generator, PDF set and order of cross-section calculations used for the normalisation
  are shown for each sample.
  The upper part of the table shows the MC samples used for the SM background coming from leptons with the same charge (MPI stands for multiple parton interactions), the lower part gives the background sources arising in the \ee\ or \emu\ channel due to electron charge misidentification.}
\label{tab:MC_cross}
\end{table}
The CT10 \cite{ct10} PDF set is used for $WZ$, $ZZ$, \Wpm\Wmp, $W\gamma$, \ttbar, and $Wt$ processes and CTEQ6L1 for others. 
The cross-sections for MPI diboson and \Wpm\Wpm\ production are calculated at leading order (LO) in QCD. For diboson samples ($WZ$, $ZZ$ and $WW$), the cross-sections are normalised to next-to-leading order (NLO) in QCD using {\sc mcfm-6.2} \cite{mcfm}. The QCD next-to-next-to-leading-order (NNLO) and next-to-next-to-leading-logarithm (NNLL) calculations are utilised for top-quark processes \cite{top1,top2,top3,top4,top5,top6,top7,top8,top9, ttbarW}.  The Drell--Yan cross-section is also calculated at NNLO in QCD by {\sc dynnlo-1.1} with MSTW2008 NNLO \cite{dynnlo,mstw}.

Some typical same-sign dilepton signals of physics beyond the SM are simulated to evaluate the efficiency and acceptance of the event selection, which are needed to set the cross-section limits (see section \ref{sec:incllimits}). Pair production of doubly charged Higgs bosons via a virtual $Z/\gamma^{*}$ exchange is generated \cite{dchmc1}. Right-handed $W$ bosons ($W_{\mathrm{R}}$) decaying to a charged lepton and a right-handed neutrino ($N_{\mathrm{R}}$) are also used \cite{wrmc}. The production of a fourth-generation heavy $b' \bar{b'}$ pair with the $b'$ quarks decaying into a $W$ boson and either a top quark or an up-type quark is considered \cite{bprimemc}. The above processes are generated using {\sc pythia}-8.165. {\sc MadGraph} is used to simulate the coloured Zee--Babu process, in which a diquark ($S_{\mathrm{DQ}}$) with charge $\pm 2/3$ or $\pm 4/3$ decays into two same-sign leptoquarks ($S_{\mathrm{LQ}}$): $pp \rightarrow S_{\mathrm{DQ}} \rightarrow S_{\mathrm{LQ}} S_{\mathrm{LQ}} \rightarrow \ell \ell q q$ \cite{cZBM}. For this process, {\sc pythia}-8.165 is utilised for the fragmentation and hadronisation. For all signal samples mentioned above, the MSTW2008LO PDF set is used and cross-sections are calculated at LO in QCD.

The background and some of the signal samples are processed using the {\sc geant}4-based \cite{g4} ATLAS detector simulation package \cite{fullsim}. Other signal samples are produced with a fast simulation \cite{fastsim} using a parameterisation of the 
calorimeter response. Additional inelastic $pp$ interactions (referred to as `pile-up'), generated with {\sc pythia}-6.426, are overlaid on the hard-scatter events to emulate the multiple $pp$ interactions
in the current and nearby bunch crossings. The distribution of the number of interactions per bunch crossing in the MC simulation is reweighted to that observed in the data. The simulated response is also corrected for differences in efficiencies, momentum scales, and momentum resolutions observed between data and simulation.

%% file: Particles.tex
The analysis makes use of muons and electrons and the basic reconstruction and identification is explained in the following. In addition, the jet reconstruction is detailed as jets misidentified as electrons are a main source of background and as electrons and muons in the vicinity of a jet are not considered in this analysis.

Jets are reconstructed in $|\eta| < 4.9$ from topological clusters \cite{topo1} formed from the energy deposits in the calorimeter, using the anti-$k_t$ algorithm \cite{antikt} with a radius parameter of $0.4$. Jets are
calibrated \cite{topo2} using an energy- and $\eta$-dependent simulation-based
calibration scheme, with in-situ corrections based on data.
The impact of multiple overlapping $pp$ interactions is accounted for using a technique that provides an event-by-event and jet-by-jet pile-up correction \cite{jetpu}. 
To reduce the effect from pile-up, for jets with $\pt < 50$~\gev, $|\eta|<2.4$,
the \pt\ of all tracks inside the jet is summed and the fraction belonging to tracks from the primary vertex is required to be larger than 0.5. The primary vertex is defined as the interaction vertex which has the highest squared-\pt\ sum of associated tracks with $\pt > 0.4$~\gev\ found in the event. At least three charged-particle tracks must be associated with this vertex.

An electron is 
formed from a cluster of cells in the electromagnetic calorimeter associated with a track in the ID. The electron \pt\ is obtained from the calorimeter energy
measurement and the direction of the associated track.
The electron 
must be within the range $|\eta| < 2.47$ and not in the transition region between the barrel and endcap calorimeters ($1.37 < |\eta| < 1.52$). In addition, a ``tight'' \cite{elepaper} set of identification criteria need to be satisfied.
One major source of background surviving these selections comes from jets misidentified as electrons.
To suppress this background, in particular at low \pt, the electrons are required to be isolated. The sum of the transverse energies 
in the electromagnetic and hadronic calorimeter cells around the electron direction in a cone of size $\Delta R = \sqrt{(\Delta \eta)^2 + (\Delta \phi)^2} = 0.2$ is required to be less than $3 \gev + (\pt^{e} - 20 \gev) \times 0.037$, where $\pt^e$ is the electron transverse momentum. The core of the electron energy depositions in the electromagnetic calorimeter is excluded and, before the cut, the sum is corrected for lateral shower leakage and pile-up from additional $pp$ collisions. 
A further isolation cut is applied using the ID information. The sum of the \pt\ of all tracks with $\pT > 0.4$~\gev\ in a cone of size $\Delta R = 0.3$ surrounding the electron track (the latter being excluded from the sum)
 is required to be less than 10\% of the electron \pt. 
 The isolation selections were optimised using electron pairs with a mass compatible with the $Z$ boson in the data, such that the application of both isolation criteria to electrons yields an efficiency that is pile-up independent and more than 99\% for electrons with $\pt > 40$~\gev. The efficiency slowly decreases with diminishing \pt\ to around 92\% at $\pt = 20$~\gev. However, these isolation selections help to suppress the background from jets misidentified as electrons, which becomes more prominent as \pt\ decreases. To further suppress leptons from hadron decays, jets in a cone of size $\Delta R=0.4$ around the electron direction are examined. Since the jet reconstruction algorithm also reconstructs electrons as jets, any jet within $\Delta R = 0.2$ of an electron is not considered to avoid double counting. The electron is rejected if there is a remaining jet in the cone with $\pt > 25 \gev\ + \pt^e \times 0.05$, where $\pt^e$ is the electron \pt. The non-constant cut value on the jet \pt\ is placed to maintain a high efficiency for very high-\pt\ electrons. 
The background arising in particular from electrons from heavy-flavour decays is reduced by requiring that the electron track is associated with the primary vertex. 
The transverse impact parameter significance is required to be
$|d_0|/\sigma(d_0) < 3$, where $d_0$ is the transverse impact parameter and $\sigma(d_0)$ is the uncertainty on the measured $d_0$. The longitudinal impact parameter $z_0$ must fulfil $|z_0 \times \sin \theta| < 1$~mm.

Muons are 
reconstructed independently in both the ID and MS. Subsequently, these two tracks are combined based on a statistical combination of the two independent measurements using the parameters of the reconstructed tracks and their covariance matrices. 
The combined track is required to be within $|\eta| < 2.5$ and the track in the ID must have at least four hits in the SCT, at least one hit in the pixel detector and one hit in the first pixel layer if an active pixel sensor is traversed. The charge measured in the ID and the MS must match as this reduces the small effect of charge misidentification 
to a negligible level. To reduce background from heavy-flavour hadron decays, each muon
is required to be isolated in the calorimeter and the ID. The calorimeter-based isolation is chosen to be $\sum \et < 3.5 \gev + (\pt^{\mu} - 20 \gev) \times 0.06$ where the $\sum \et$ is calculated in a cone of size $\Delta R=0.3$ and $\pt^{\mu}$ is the muon transverse momentum. The ID-based isolation is defined as $(\sum \pt) / \pt^{\mu} < 0.07$, where the sum runs over ID tracks with $\pt > 1$~\gev\ in a cone of size $\Delta R = 0.3$ surrounding the muon track, the latter being excluded from the sum. These isolation cuts result in an efficiency for muons from $Z$ decays in data which is above 99\% for muons with $\pt^{\mu} > 40$~\gev\ and decreases to around 90\% at $\pt^{\mu} = 20$~\gev. 
As for electrons, with diminishing \pt\ the background becomes more pronounced and these cuts result in a better background rejection. To further reduce the background a nearby jet veto is applied, similar to that for electrons: a muon is rejected if a jet with $\pt > 25 \gev\ + \pt^{\mu} \times 0.05$ is found in a cone of size $\Delta R = 0.4$ around the muon. 
The muons are required to be associated with the primary vertex by requiring $|d_0|/\sigma(d_0) < 3$, $|d_0| < 0.2$~mm and $|z_0 \times \sin \theta| < 1$~mm.

%% file: Selection.tex
This analysis uses the 2012 $pp$ collision data collected at a centre-of-mass energy of 8~\tev\ with, on average, 21 interactions per bunch crossing. After requiring that all detector components are operational, the dataset amounts to 20.3 \ifb\ of integrated luminosity.

The events are selected by electron and muon triggers. Events in the \mumu\ channel are selected by a dimuon trigger, which requires one muon with transverse momentum larger than 18~\gev\ and another muon with $\pt > 8$~\gev.
At the trigger level, muons are identified by requiring that the candidate muon tracks are reconstructed in both the MS and the ID. 
Dielectron events are recorded if the event contains two electrons with a \pt\ larger than 12~\gev\ that satisfy the ``loose'' identification criteria.
In the \emu\ channel 
events are selected by the trigger if both an electron and a muon ($e\mu$) are found or if a high-\pt\ electron is identified. 
For the $e \mu$ trigger the electron must have $\pt>12$ \gev\ satisfying the ``medium'' set of identification criteria, whereas the muon must have $\pt>8$ \gev. The high-\pt\ electron trigger selects events containing electrons, which satisfy the ``medium'' identification criteria and have $\pt>60$ \gev. These triggers yield a sample of dilepton events with high efficiency over the whole $\pt$ range considered in this analysis. 

The selected events must have a reconstructed primary vertex and contain lepton pairs with $\pt > 25$~\gev\ for the leading lepton and $\pt > 20$~\gev\ for the subleading one. These leptons must have the same electric charge, meet the above selection requirements, and have an invariant mass $m(\ell \ell') > 15$~\gev. To reduce the background from leptons from $Z$ boson decays, events in which an opposite-sign, same-flavour lepton pair is found to be consistent with the invariant mass of the $Z$ boson ($|m_{\ell \ell} - m_{Z}| < 10$~\gev) are rejected.
In the \ee\ channel, electron pairs in the mass range between 70~\gev\ and 110~\gev\ are vetoed as this region is used for the background estimates (see section \ref{sec:oppsignbkg}).
Any combination of two leptons with the same charge and with $\pt>25$~\gev\ and $\pt>20$~\gev\ respectively is included. This allows more than one lepton pair per event to be considered, which happens in fewer than 0.1\% of the events.

%% file: Background.tex
The backgrounds in this search can be subdivided into prompt background, backgrounds from SM processes with two opposite-sign leptons where the charge of one of the leptons is misidentified and non-prompt background. 
Prompt leptons originate from a decay of a $W$ boson, $Z$ boson, and include any leptonic products of a prompt $\tau$ lepton decay. Non-prompt leptons are from decays of long-lived particles and mainly arise from semileptonic decays of heavy-flavour hadrons (containing $b$ or $c$-quarks). 
Hadrons or overlapping hadrons within a jet which may be misidentified as an electron are also called non-prompt leptons in the following.
The prompt background comes from SM processes producing two same-sign leptons from the primary vertex, and arises mainly from $WZ, ZZ$, \Wpm \Wpm, \ttbar $W$, and \ttbar $Z$ production (see section \ref{sec:promptbkg}). The method used to estimate the background from lepton charge misidentification is described in section \ref{sec:oppsignbkg}.
Background from non-prompt leptons
can arise from various sources and is discussed in section \ref{sec:nonpromptbkg}. For electrons, the main sources are jets misidentified as electrons 
and semileptonic decays of hadrons containing $b$- or $c$-quarks. For muons, the main contribution arises from semileptonic decays of heavy-flavour hadrons. Small contributions also come from pions and kaons that decay in flight and misidentified muons from hadronic showers in the calorimeter which reach
the MS and are incorrectly matched to a reconstructed ID track.

\subsection{Background from prompt same-sign lepton pairs}
\label{sec:promptbkg}
The background from SM processes in which prompt same-sign lepton pairs are produced is determined from MC simulations. Processes other than those listed in table \ref{tab:MC_cross} 
do not contribute significantly to this type of background and are neglected.
In all of these samples, only reconstructed leptons are 
considered that are matched to a lepton at generator level from a decay of a 
$W$ boson, a $Z$ boson, and include any leptonic products of a prompt $\tau$ 
lepton decay. Leptons from any other sources are discarded to avoid double counting with the background from non-prompt leptons. 

\subsection{Background from opposite-sign lepton pairs}
\label{sec:oppsignbkg}
Monte Carlo samples are also used to simulate the contributions from processes in which opposite-sign lepton pairs are produced and one of the leptons has an incorrect charge assigned. 
In principle, this charge misidentification can occur for muons as well as electrons. However, a study using muons from $Z$ boson decays shows that
this effect is negligible in this analysis.
In the case of electrons, the dominant process that leads to charge misidentification is electrons emitting hard bremsstrahlung and subsequently producing electron--positron pairs by photon conversion, with one of these leptons having a high \pt. Typically these conversions would be reconstructed as such, but in some asymmetric conversions only one of the tracks is reconstructed and the charge may be opposite to the charge of the original lepton that radiated the photon. 
The charge misidentification probability is measured using electrons from $Z$ boson decays. This is done in a data-driven way using the same likelihood method as used in ref.~\cite{2011SS}. The electrons are required to pass the same selection 
cuts as detailed in sections \ref{sec:particles} and \ref{sec:selection}, and are selected by requiring same-sign electron pairs with an invariant mass between 80 \gev\ and 100 \gev. This results in a very pure sample of electron pairs for which the charge of one of the electrons is incorrectly assigned. A comparison  between data and MC events shows that the charge misidentification rate as a function of \et\ is well modelled in the simulation. 
Simulation is used to predict the backgrounds from Drell--Yan, \ttbar, and \Wpm \Wmp\ production, correcting the event weights for events with a
charge-misidentified electron by an $|\eta|$-dependent factor derived from these studies.
The process $V\gamma \rightarrow \ell \ell' \gamma \rightarrow \ell \ell' ee, V = W, Z$ can also give rise to same-sign lepton pairs when the photon converts. Since this background is closely related to the electron charge misidentification, the same correction factor is applied to the electrons from converted photons in the MC simulation. 
The contribution of conversions from $Z\gamma$ events is implicitly accounted for in the simulation of the $Z/\gamma^∗$ process and is included in the charge misidentification category.

The uncertainty in the measurement of the charge misidentification rate is estimated by varying the invariant mass window and by loosening the isolation criteria. The total systematic error varies between 6\% and 20\% depending on the pseudorapidity. For tracks with very high \pt, 
the charge can be incorrectly assigned due to the imperfect resolution and alignment of the detector, since the curvature of the tracks is very small.
An additional uncertainty of 20\% is assigned to the misidentification rate for electrons with $\pt > 100$~\gev,  
based on the following study. 
Since the charge misidentification rate is affected by the detector material description in the simulation, simulations with different material descriptions\footnote{These simulations contain the following additional material: 5\% in the whole of the ID, 20\% for the pixel and SCT services each, 15\% $X_0$ at the end of the SCT/TRT endcap and 15\% $X_0$ at the ID endplate.} are compared with the nominal simulation. 
The largest variation is found in the endcap region and this difference is taken as the overall uncertainty.

\subsection{Background from non-prompt leptons}
\label{sec:nonpromptbkg}
The background from non-prompt and misidentified leptons is determined in a data-driven way as a function of the lepton \pt\ and $|\eta|$. For both the electrons and the muons (see sections \ref{sec:fakemuon} and {\ref{sec:fakeelectron}) the following method is used to predict the contribution of non-prompt leptons in the signal region. A background region is defined that contains predominantly non-prompt leptons or jets that are kinematically similar to those in the signal region. A factor $f$ is determined, which is the ratio of the number of leptons satisfying a given selection criterion ($N_{\mathrm P}$) to the number of leptons, which do not meet this requirement but satisfy a less stringent criterion ($N_{\mathrm F}$),
\begin{equation}
f = \frac{N_{\mathrm{P}} - N_{\mathrm{P}}^{\mathrm{prompt}}}{N_{\mathrm{F}}  - N_{\mathrm{F}}^{\mathrm{prompt}}}.
\label{eq:fakefactor}
\end{equation}
This ratio, which is determined as a function of \pt\ and $|\eta|$, is corrected for the residual contribution of prompt leptons ($N_{\mathrm{P}}^{\mathrm{prompt}}$ and $N_{\mathrm{F}}^{\mathrm{prompt}}$) using MC simulations. 
The factors $f$ are calculated separately for the different pass and fail criteria that apply to the signal and validation regions discussed in sections \ref{sec:particles} and \ref{sec:validation}, respectively. 

The total number of events with non-prompt leptons, $N_{\mathrm{NP}}$, in a given signal or validation region is predicted to be
\begin{equation}
N_{\mathrm{NP}} = \sum_{i}^{\mathrm{N_{P_l+F_s}}} f_{\mathrm{s}}(p_{\mathrm{Ti}},|\eta_{i}|) + \sum_{i}^{\mathrm{N_{F_l+P_s}}} f_{\mathrm{l}}(p_{\mathrm{T}i},|\eta_{i}|) - \sum_{i}^{\mathrm{N_{F_l+F_s}}} f_{\mathrm{l}}(p_{\mathrm{T}i},|\eta_{i}|) \times f_{\mathrm{s}}(p_{\mathrm{T}i},|\eta_{i}|).
\label{eq:fake_pred}
\end{equation}
The first term is the number of pairs $N_{\mathrm{P_l+F_s}}$, where the leading lepton (denoted by l) fulfils the selection requirements ($\mathrm{P_l}$) and the subleading lepton (denoted by s) fails to satisfy its selection criteria ($\mathrm{F_s}$). This is weighted per pair by the factor $f_\mathrm{s}$ of the subleading lepton (the lepton which failed). Similarly for the second term, the leading lepton fails its selection ($\mathrm{F_l}$) and the subleading lepton satisfies its respective selection criteria ($\mathrm{P_s}$), hence the weight per pair is given by the factor $f_\mathrm{l}$ for the leading lepton. The last term is included to avoid double counting and represents the case where both the leading and subleading leptons fail to satisfy their respective criteria, and so a weight for each lepton is needed.
The factors $f_\mathrm{l}$ and $f_\mathrm{s}$ are taken to be the same in regions where both leptons fulfil the same selection requirement, as in the signal region and some of the validation regions.

\subsubsection{\boldmath Measurement of $f$ for muons}
\label{sec:fakemuon}
In the case of muons the factor $f$ is determined using a background region that contains mainly muons from semileptonic decays of $b$- and $c$-hadrons. 
This region is defined by taking advantage of the long lifetimes of $b$- and $c$-hadrons. Events are selected containing same-sign muon pairs that fulfil the same selection criteria as for the signal region (see section \ref{sec:particles}), but requiring that at least one of the leptons has $|d_0|/\sigma(d_0) > 3$ and $|d_0| < 10$~mm. The same dimuon trigger as in the signal region is used here. The number of muons passing these impact parameter cuts is subdivided in two categories: $N_{\mathrm{P}}$, which are those muons passing the calorimeter and track-based isolation cuts, and $N_{\mathrm{F}}$, which contain those muons which fail the calorimeter-based isolation cut, the track-based isolation cut or both. 
The measured factor $f$ is between 0.11 and 0.20 for muons with $\pt > 20$~\gev.

Muons from $b$- and $c$-hadron decays tend to have large impact parameters, to be accompanied by other tracks, and to be 
less isolated than prompt muons, which are associated with the primary vertex.
Since the muon isolation can depend on the impact parameter, a correction needs to be applied to the factor $f$ in the signal region.
This correction 
is determined using $b \bar{b}$ and $c \bar{c}$ MC simulations. 
In both the signal region and background region $f$ is determined and the correction is then given by the ratio of these two quantities.
As the correction factor is found to be independent of \pt, the overall value of 1.3 is measured using same-sign muon pairs with $\pt > 20$~\gev\ and $m_{\mu \mu} > 15$ \gev.

The two main sources of uncertainty in this procedure come from the uncertainty associated with the correction made to $f$ before its use in the signal region, which comes primarily from the statistical error on the MC sample used in its derivation, and from the statistical uncertainties in the background data sample.
Further sources of uncertainty arise from the prompt background subtraction and a possible difference between the signal and background region in the fraction of non-prompt muons from heavy-flavour decays and light particles, such as pions and kaons, which decay in flight. The total uncertainty on $f$ is 17\% at $\pt \approx 20$ \gev\ increasing to 23\% for $\pt \approx 60$ \gev. A value of 100\% is used for $\pt > 100$ \gev\ due to a lack of statistics to determine $f$.

\subsubsection{\boldmath Measurement of $f$ for electrons}
\label{sec:fakeelectron}
To measure $f$ for electrons a dijet data sample is selected which
contains events with either a jet misidentified as an electron or a non-prompt electron from a semileptonic decay of $b$- and $c$-hadrons.
The selected region consists of events that contain exactly one electron 
with $\pt > 20$ \gev\ and a jet in the opposite azimuthal direction ($\Delta \phi(e, jet) > 2.4$). The electron has to satisfy the ``medium'' identification criteria, the same impact parameter cuts as for the signal region and is rejected if, after removal of any jet within $\Delta R = 0.2$ of the electron,
there is a remaining jet within  $\Delta R=0.4$. These events are selected by a set of prescaled single-electron triggers with different electron $\pt$ thresholds. 
To ensure that the electron and the jet are well balanced in terms of energy, a $\pt > 30$ \gev\ requirement is applied to the jet. The different cut value from the electron case accounts for the differences in the electron and jet energy scale calibrations and for energy depositions from other decay products in the isolation cone around the electron direction. 
Electron pairs from $Z/\gamma^*$ or \ttbar\ events do not satisfy the above selection criteria. 
To suppress electrons from $W$ boson decays, events are rejected if the transverse mass\footnote{Transverse mass $m_{\mathrm{T}} = \sqrt{2 \times \et^{\mathrm{\ell}} \times \et^{\mathrm{miss}} \times (1 - \cos \Delta \phi)}$, where $\Delta \phi$ is the azimuthal angle between the directions of the electron and the missing transverse momentum (with magnitude $\et^{\mathrm{miss}}$). The missing transverse momentum is defined as the momentum imbalance in the plane transverse to the beam axis and is obtained from the negative vector sum of the momenta of all particles detected in the event \cite{misset}.} exceeds 40 \gev.

The number $N_{\mathrm{P}}$ is calculated from events in this background region for which the electron satisfies the same electron selection criteria as applied in the signal region. The value of $N_{\mathrm{F}}$ is based on electron candidates satisfying the signal selection criteria but passing less stringent electron identification cuts (``medium'')
and failing to meet the calorimeter-based or track-based isolation requirements, or both. The numbers are corrected for the small remaining contribution from prompt electrons (see equation (\ref{eq:fakefactor})).
The measured factor $f$ is 0.18 at $\pt = 20$~\gev\ and increases to around 0.3 for $\pt \approx 100$~\gev. 
The main systematic uncertainty is due to the jet requirements in the event selection. This effect is estimated by varying the jet \pt\ between 30 \gev\ and 50 \gev, which leads to an uncertainty ranging between 10\% and 30\% depending on the electron \pt. 
Other systematic uncertainties arise from a possible difference in the heavy-flavour fraction in the signal and background region, and the prompt background subtraction.   
The total uncertainty varies between approximately 40\% at $\pt \approx 20$ \gev\ and 13\% for $\pt \approx 100$ \gev. 
Due to a lack of statistics to calculate $f$ for electrons with $\pt > 100$~\gev, the value of $f$ for $60 < \pt <100$~\gev\ electrons is used, and the uncertainty is increased to 100\%.

\subsection{Validation of background extraction methods}
\label{sec:validation}
The background predictions from the various sources (prompt, non-prompt and charge misidentification) are validated using different methods, as discussed in the following and summarised in table \ref{tab:val_regions}. 

\begin{table}[htbp]
\begin{center}
\begin{tabular}{ l | p{0.5\textwidth} }
\hline
Validation method & Primary background or validation criterion \\
\hline
Weak isolation VR's & Electron and muon non-prompt background \\
Fail-$d_0$ VR's & Electron and muon non-prompt background \\
Medium VR & Electron and muon non-prompt background \\
Low muon \pt\ VR & Muon non-prompt background \\
Opposite-sign VR & Normalisation, efficiencies, lepton $p_T$ scale and resolution \\
Prompt VR & Prompt MC background predictions \\
Same-sign dielectron $Z$ peak closure test & Charge misidentification correction applied to opposite-sign MC background samples\\
\hline
\end{tabular}
\end{center}
\caption{A summary of the validation methods used and an explanation of the type of background the methods are testing or which data-driven estimates they validate. These tests are carried out using validation regions (VR) or closure tests and are discussed in detail in the text.}
\label{tab:val_regions}
\end{table}
To test the predictions for the non-prompt background, validation regions (VR) that contain same-sign lepton pairs are defined.
In these regions one or both of the leptons fail one of the signal selection cuts but pass a less stringent cut, which is called a ``weaker'' selection in the following.
The dilepton invariant mass, the lepton \pt\ and $\eta$ distributions are compared between data and predictions. 
\begin{figure}[hbtp]
\begin{center}
\subfigure[]{
	\includegraphics[width=0.47\columnwidth]{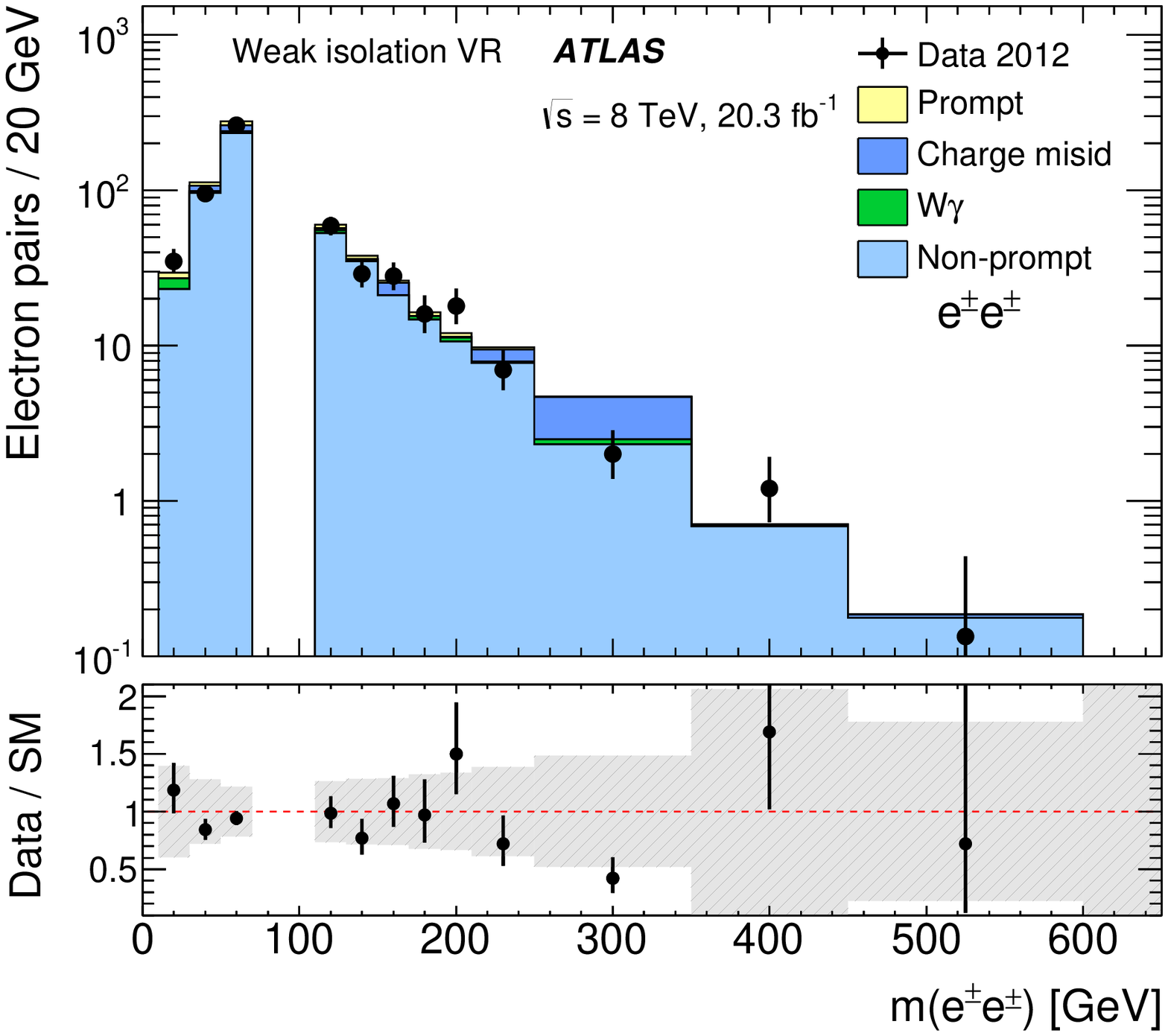}
}
\subfigure[] {
  \includegraphics[width=0.47\columnwidth]{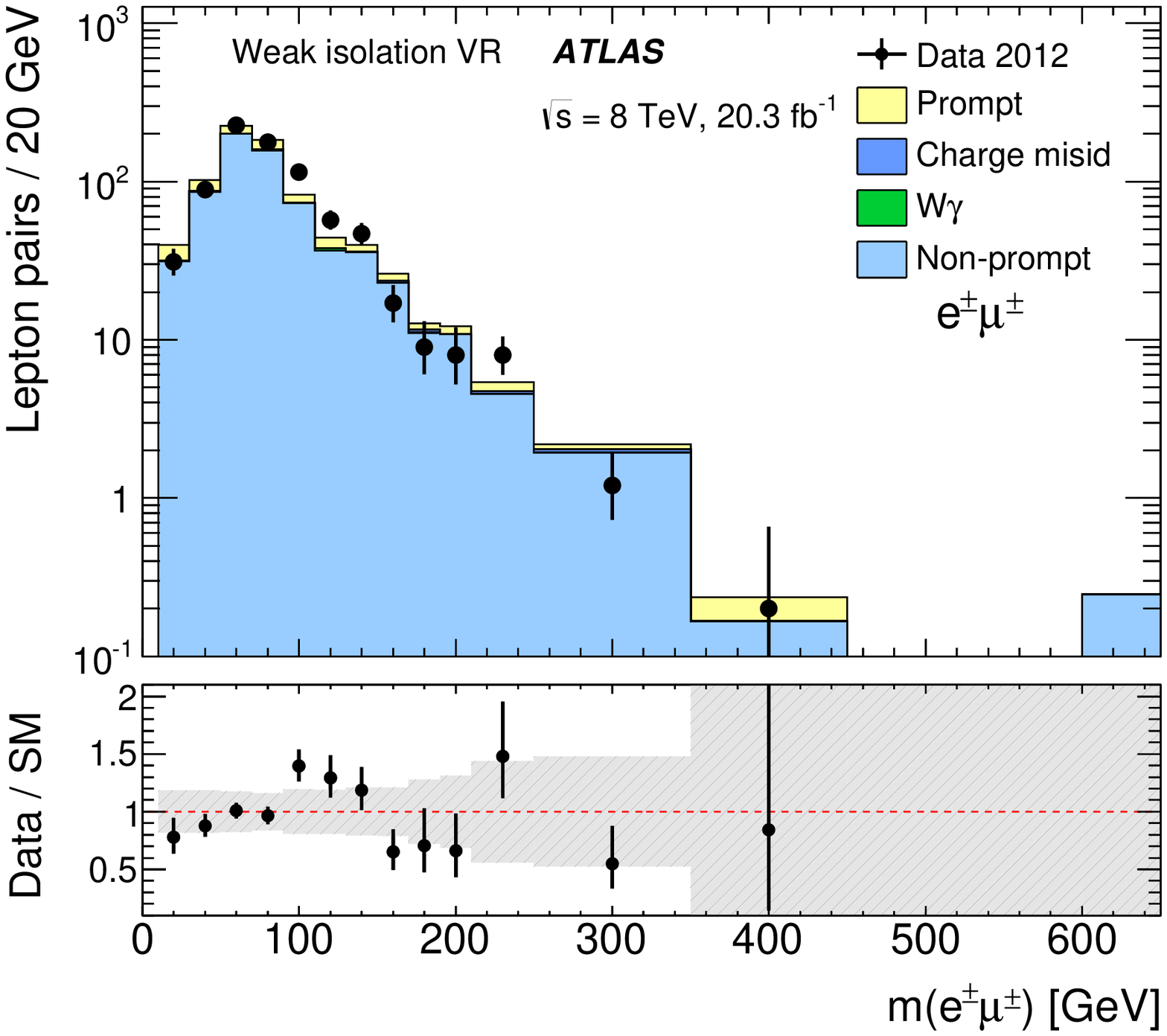}
}
\subfigure[] {
  \includegraphics[width=0.47\columnwidth]{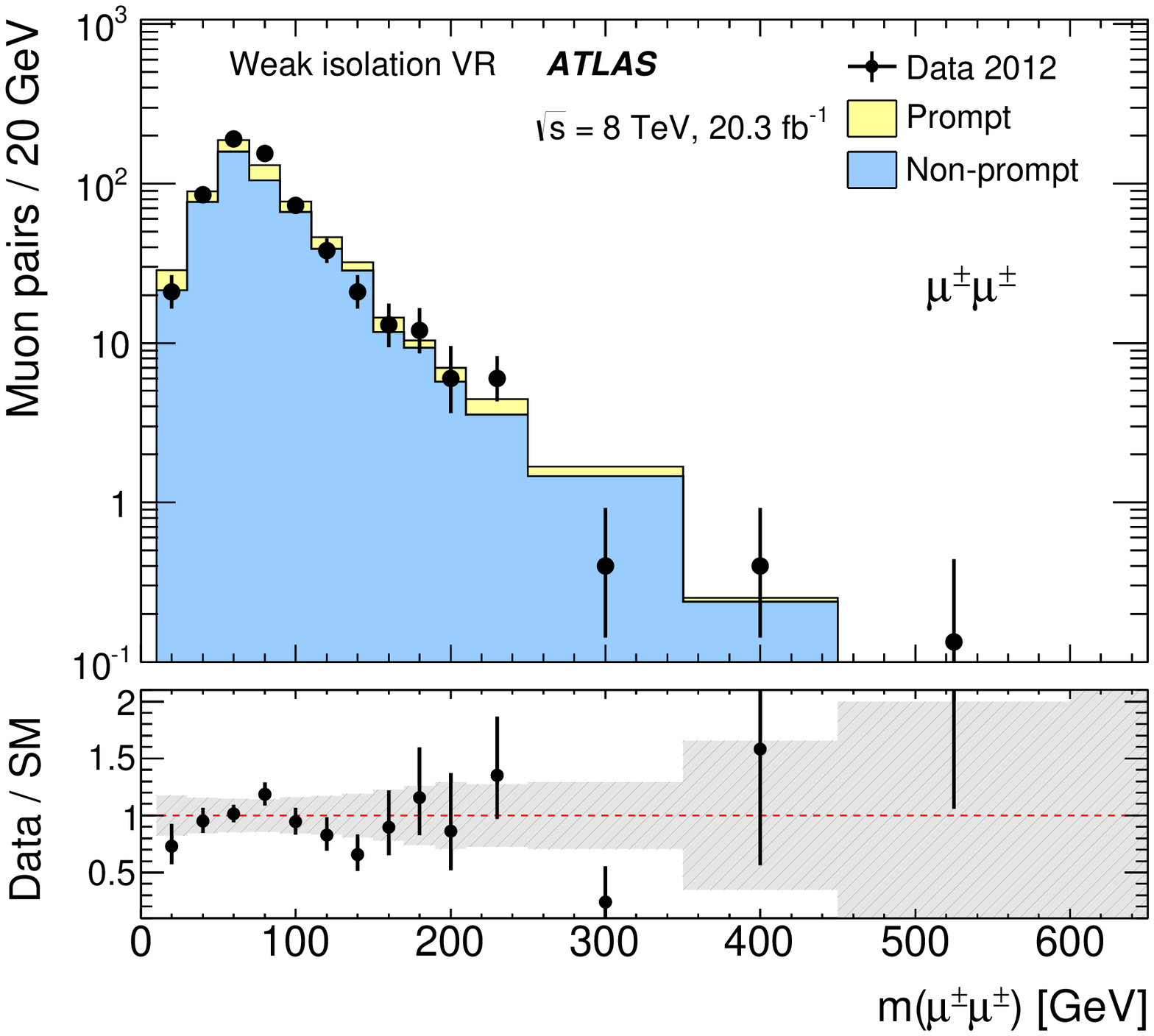}
}
\subfigure[] {
\includegraphics[width=0.47\columnwidth]{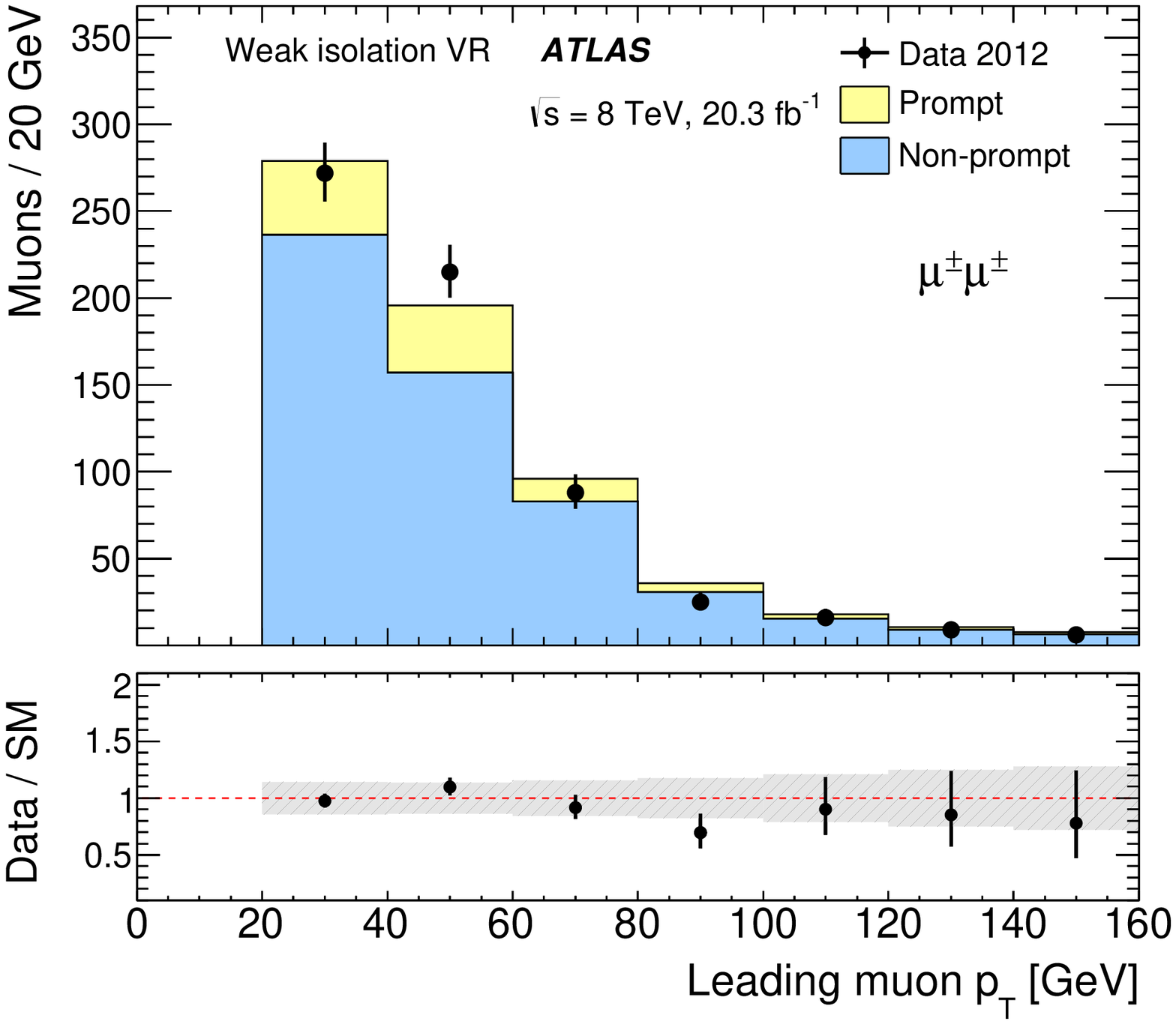}
}
\caption{\label{fig:check1} Invariant mass distributions in one of the validation regions (VR) used in the (a) \ee\ (b) \emu\, and (c) \mumu\ channels. The \pt\ distribution for the leading muon in the \mumu\ channel is shown in (d). In (a), (c) and (d) the leading lepton of the pair passes the signal isolation cuts while the subleading lepton passes the ``weak'' isolation cuts (weak isolation on subleading lepton VR). The mass range between 70~\gev\ and 110~\gev\ is not included in the \ee\ channel as this region is used to estimate the background from charge misidentification.
The electron passes the isolation cuts defined for the signal region while the muon passes the ``weak'' isolation cuts in (b) (weak isolation on muon VR). 
The data are compared to the background expectations and the lower panels show the ratio of data to the background prediction. 
The error bars on the data points show the statistical uncertainty and the dashed band shows the total uncertainties of the predictions. The last bin in the histograms includes overflows, and is normalised as though it is 50 \gev\ wide in (a) - (c) and 20 \gev\ wide in (d).}
\end{center}
\end{figure}
\begin{table}[htbp]
    \begin{center}
    \begin{tabular}{ l | c c c }
        \hline
 & \multicolumn{3}{c}{Number of electron pairs} \\
        Region 		& Predictions		& Data		&  Difference/$\sigma$ \\
            \hline
        Weak isolation on both leptons VR	& $ 280 \pm 130 $& $285$	& $\hspace*{0.3cm}0.0$\\
        Weak isolation on leading lepton VR&  $ 190 \pm 60 $		& $ 224$&   $-0.6$ \\
        Weak isolation on subleading lepton VR&  $ 620 \pm 120 $		& $574$ &   $+0.4$ \\
        Medium VR	& $ 195 \pm 32 $& $ 217 $&  $-0.7$\\
            Opposite-sign VR & $4740000 \pm 330000$ & $4895830$ & $-0.5$\\
Prompt VR with $m_{\ell^{\pm} \ell^{\pm}} > 15$ \gev & $275 \pm 23$ & $268 $ & $+0.3$\\ 
       $Z$ peak closure test & $12700 \pm 1300$ & $11793$ & $+0.7$\\
\hline\hline
 & \multicolumn{3}{c}{Number of electron--muon pairs} \\
        Region 		& Predictions		& Data		&  Difference/$\sigma$ \\
\hline
        Weak isolation on muon VR & $790 \pm 130$ &      $800$ &    $-0.1$ \\
        Weak isolation on electron VR & $750 \pm 150$ &      $965 $ &  $-1.4$ \\
        Fail-$d_0$ VR    & $249 \pm 19$ &       $216 $ &   $+1.7$ \\
        Low muon \pt\ VR  &  $211 \pm 12$ &        $201 $ &    $+0.8$ \\
Opposite-sign VR & $70400 \pm 4700$ & $71771 $ & $-0.3$\\
Prompt VR with $m_{\ell^{\pm} \ell^{\pm}} > 15$ \gev & $950 \pm 60$ & $1001$ & $-0.8$\\ 
       \hline \hline
 & \multicolumn{3}{c}{Number of muon pairs} \\
        Region 		& Predictions		& Data		&  Difference/$\sigma$ \\
\hline
        Weak isolation on both leptons VR & $280 \pm 40$ &        $283$ &  $-0.1$ \\ 
        Weak isolation on leading lepton VR & $199 \pm 25$ &        $199$ &   $\ \hspace*{0.15cm}0.0$ \\ 
       Weak isolation on subleading lepton VR & $697 \pm 90$ &        $652$ &  $+0.5$ \\ 
        Fail-$d_0$ VR  &  $250 \pm 31$ &        $255$ &  $-0.2$ \\  
Opposite-sign VR & $8144000 \pm 10000$ & $8216983$ & $-0.7$\\
Prompt VR with $m_{\ell^{\pm} \ell^{\pm}} > 15$ \gev & $651 \pm 43$ & $714$ & $-1.5$\\ 
        \hline
    \end{tabular}
    \end{center}
\caption{Expected and observed numbers of lepton pairs for the different 
validation regions, explained in detail in the text. 
The uncertainties on the predictions include the statistical and systematic uncertainties. The column 'Difference/$\sigma$' is calculated by dividing the difference between the predictions and the data by the uncertainty ($\sigma$) of the prediction.
}
\label{tab:agreement}
\end{table}

One of the validation regions selects
a leading lepton that satisfies the signal selection criteria and a subleading lepton that passes the ``weak'' isolation cuts, which means the subleading lepton fails to meet the signal calorimeter or ID-based isolation requirement and instead passes isolation cuts that are loosened by 4 \gev\ (weak isolation on subleading lepton VR). 
For this region, the factor $f$ for the subleading lepton 
is determined according to equation (\ref{eq:fakefactor}) using as pass criteria the ``weak'' isolation requirements and as fail criteria the loose isolation criteria applied in the ``weak'' selection. 
The invariant mass distributions for the three final states in this validation region are shown in figure \ref{fig:check1} together with one example for the \pt\ distribution.  
The predictions agree well with the data.

In another validation region the ``weak'' isolation selection is applied to the leading lepton (weak isolation on leading lepton VR). 
In the \mumu\ and \emu\ channel one validation region 
(fail-$d_0$ VR) requires that one muon has an impact parameter significance of $|d_0|/\sigma(d_0) > 3$ and in order to increase the statistics the $|d_0|$ cut is loosened to 10~mm. In the \ee\ final state one region (medium VR) contains same-sign electron pairs, in which one of the electrons fails the ``tight'' identification cuts but passes the looser ``medium'' selections instead. One region (low muon \pt\ VR) used in the \emu\ channel selects same-sign electron--muon pairs, where both leptons satisfy the signal selection criteria but the muon has a transverse momentum between 18 \gev\ and 20 \gev. 

To test the trigger and reconstruction efficiencies, as well as the lepton momentum scale and resolution, an opposite-sign validation region (opposite-sign VR) is defined. 
This region is populated with prompt lepton pairs that pass the same event selection as for the same-sign signal region, but the leptons have opposite charge. 
In order to cross-check the normalisation of the dominant $WZ$ and $ZZ$ MC predictions, a prompt validation region (prompt VR) is utilised. Events are selected in which at least three leptons are present. One pair must be from a same-sign lepton pair and another from an opposite-sign same-flavour lepton pair that has an invariant mass ($m_{\ell \ell}$) compatible with the $Z$ boson mass ($|m_{\ell \ell} - m_{\mathrm{Z}}| < 10$ \gev). The data and predictions are compared for different cuts on the invariant mass of the same-sign lepton pair. To test the correction factor for the charge misidentification (see section \ref{sec:oppsignbkg}), the factor $f$ is applied to simulated
$Z$ decays into an electron pair where one electron is reconstructed with the wrong charge.
A closure test is carried out in the region around the $Z$ peak. This test shows that the shape of the background from charge misidentification is correctly reproduced. 

In all channels and validation regions the agreement between observation and prediction is good, as can be seen in table \ref{tab:agreement}. 
The agreement between data and predictions is typically better than $1\sigma$ and at most $1.7\sigma$.

%% file: Uncertainties.tex
\begin{table}[ht]
\begin{center}
\begin{tabular}{l|l|c|c|c}
\hline
Source & Process & \multicolumn{3}{c}{Uncertainty} \\
& & \ee & \emu & \mumu\\
\hline
\multirow{2}{*}{Trigger} & Signal and background & \multirow{2}{*}{2.1-2.6\%} & \multirow{2}{*}{2.1-2.6\%} & \multirow{2}{*}{2.1--2.6\%} \\
& from MC simulations & & &\\
\hline
Electron reconstruction & Signal, prompt &\multirow{2}{*}{1.9--2.7\%} & \multirow{2}{*}{1.4\%}& \multirow{2}{*}{}\\
and identification & background&&&\\\hline
Muon reconstruction & Signal, prompt & \multirow{2}{*}{} & \multirow{2}{*}{0.28\%}& \multirow{2}{*}{0.6\%}\\
and identification & background&&&\\\hline
Electron charge  & Opposite-sign& \multirow{2}{*}{9\%} & \multirow{2}{*}{1.2\%}  & \multirow{2}{*}{}\\
misidentification& backgrounds&&&\\\hline
Determination of & Non-prompt &\multirow{2}{*}{22\%}&\multirow{2}{*}{24\%}&\multirow{2}{*}{17\%}\\
factor $f$ for $e/\mu$& backgrounds& & & \\\hline
\multirow{2}{*}{Luminosity} & Signal and background& \multirow{2}{*}{2.8\%}& \multirow{2}{*}{2.8\%}& \multirow{2}{*}{2.8\%}\\
& from MC simulations&&&\\\hline
\multirow{2}{*}{MC statistics} & Backgrounds from &  \multirow{2}{*}{5\%}&  \multirow{2}{*}{1.6\%}&  \multirow{2}{*}{1.3\%}\\
& MC simulations &&&\\\hline
Photon misidentification & \multirow{2}{*}{$W\gamma$} & \multirow{2}{*}{13\%}& \multirow{2}{*}{11\%}& \multirow{2}{*}{}\\
as electron&&&&\\\hline
\multirow{2}{*}{MC cross-sections} & Prompt, opposite-& \multirow{2}{*}{4\%}& \multirow{2}{*}{2.5\%}& \multirow{2}{*}{4\%} \\
& sign backgrounds & &&\\
\hline
\end{tabular}
\end{center}
\caption{Sources of systematic uncertainty (in \%) on the signal yield and the expected background predictions, described in the second column, for the mass range $m_{\ell \ell} > 15$ \gev.
}
\label{tab:syst}
\end{table}
The systematic uncertainties considered in this analysis are summarised in table \ref{tab:syst}. 
Experimental systematic uncertainties arise from the trigger selection and the lepton reconstruction and identification. These include the effects of the energy scale and resolution uncertainties.
Also shown is the overall uncertainty in the \ee\ and \emu\ channels from electron charge misidentification (discussed in section~\ref{sec:oppsignbkg}) and the non-prompt background estimation (presented in section~\ref{sec:nonpromptbkg}). 

The uncertainty on the integrated luminosity is 2.8\%. It is derived following the same methodology as that detailed in ref.~\cite{lumipaper}.
Another systematic uncertainty is due to the limited number of events available in the MC samples and also the data control samples used for the background predictions. The overall effect from the MC samples used per channel is shown in table  \ref{tab:syst}. 
Systematic uncertainties on different physics processes from the same source are assumed to be 100\% correlated.
An example is the charge misidentification rate uncertainty for the $Z/\gamma^{*}$, $t\bar{t}$, $WW$ and $W\gamma$ samples.
 
Theoretical uncertainties
on the production cross-section arise from the choice
of renormalisation and factorisation scales in the fixed-order
calculations as well as the uncertainties on the PDF sets and the
value of the strong coupling constant $\alpha_{\mathrm{s}}$ used in the perturbative expansion. The uncertainties due to the renormalisation and factorisation scales are found by varying the scales by a factor of two 
relative to their nominal values. The PDF and $\alpha_{\mathrm{s}}$ uncertainties are determined using different PDF sets and PDF error sets following the recommendations documented in ref.~\cite{pdf4lhc}. 
The uncertainties on the MC modelling of background processes are estimated by testing different generators as well as parton shower and hadronisation models. 
The resulting total cross-section uncertainties are 7\% for $WZ$~\cite{diboson}, 5\% for $ZZ$~\cite{diboson}, and 22\% for $t\bar{t}V$~\cite{ttbarW,uncertttz}. The uncertainties on \Wpm\Wpm\ cross-sections and diboson production in MPI processes are taken to be 50\% and 100\% respectively, but their contributions to the final results are small.

%% file: Results.tex
\subsection{Signal region}
\begin{figure}[thbp]
\begin{center}
\subfigure[]{\includegraphics[width=0.47\columnwidth]{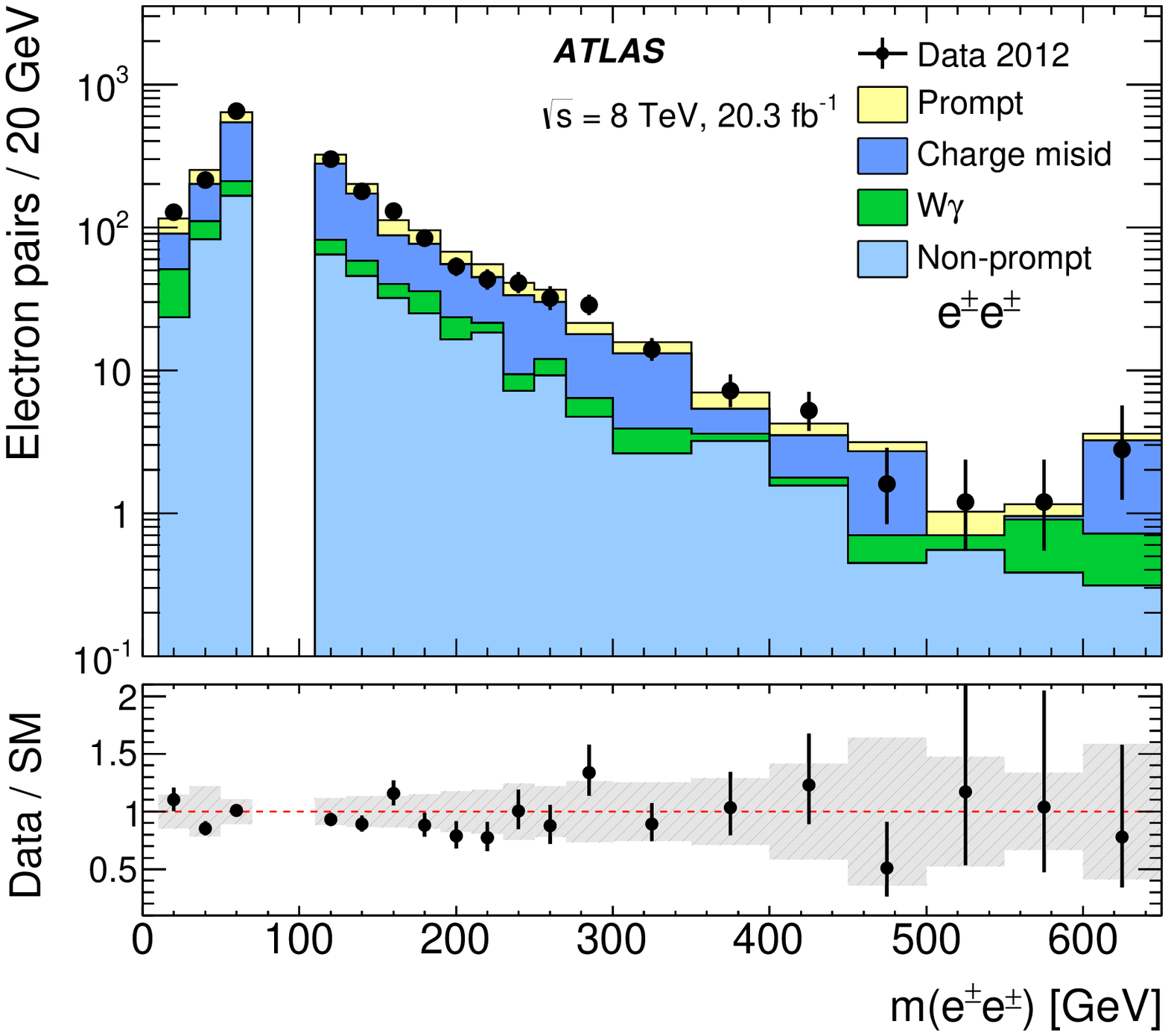} }
\subfigure[]{\includegraphics[width=0.47\columnwidth]{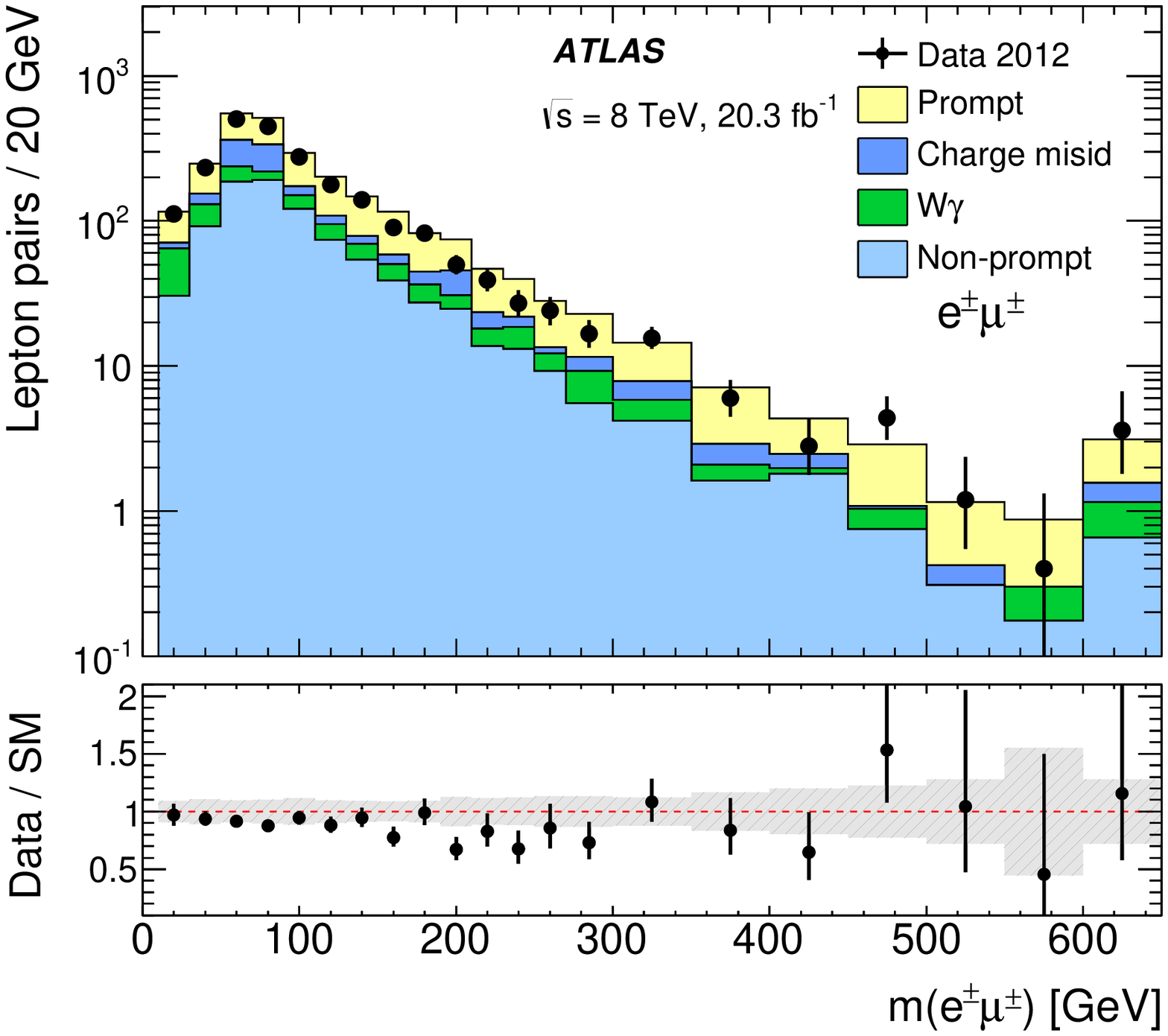} }
\subfigure[]{\includegraphics[width=0.47\columnwidth]{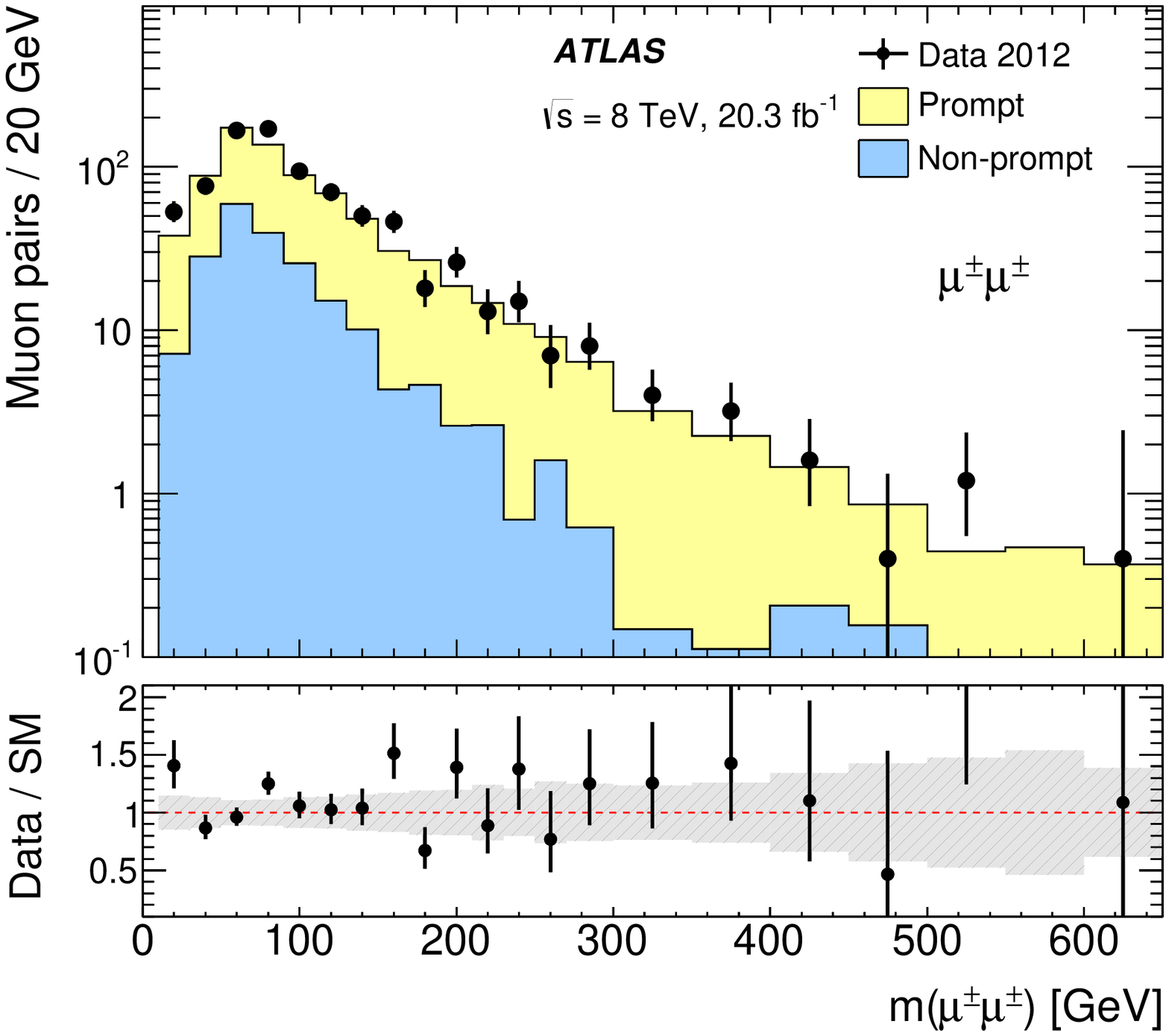} }
\caption{\label{fig:dileptonmass} Invariant mass distribution of (a) \ee\ (b) \emu\ and (c) \mumu\ pairs as a function of a threshold on the dilepton mass in the same-sign signal region. The mass range between 70~\gev\ and 110~\gev\ is not included in the \ee\ channel as this region is used to estimate the background from charge misidentification. The data are compared to the SM expectations and the lower panels show the ratio of data to the background prediction.
  The error bars on the data points show the statistical uncertainty and the dashed band shows the total uncertainties of the predictions. The last bin in the histograms includes overflows, and is normalised as though it is 50 \gev\ wide.
}
\end{center}
\end{figure}

The invariant mass distributions for the data and the expected SM background are shown in 
figure \ref{fig:dileptonmass}, separately for the \ee, \emu\ and \mumu\ final states.
In general, good agreement is seen in both the total normalisation and shapes for all channels within the uncertainties. The last bin in the figures contains the overflow bin. There is no event in the overflow in the \mumu\ channel, while the mass distribution extends up to around 1300 (1100) \gev\ in the \ee\ (\emu) channel. 
The expected and observed numbers of events for several cuts on the dilepton mass for each final state are given in table \ref{tab:SMbackground}, which also shows the contributions from the different background types. 
In the \ee\ channel the dominant background contribution comes from charge misidentification of electrons from the Drell--Yan process. In the \emu\ and \mumu\ channel the prompt production dominates the background. The prompt background predominantly arises from $WZ$ boson production, which amounts to around 70\% of the prompt background. This fraction slightly decreases for high-mass dilepton pairs. Other contributions are from $ZZ$ and \Wpm\Wpm\ production and a very small fraction comes from the $t\bar{t}W$ and $t\bar{t}Z$ production or from diboson production in MPI processes. For dilepton masses $m_{\ell \ell} > 500$ \gev, the contribution from \Wpm\Wpm\ and $ZZ$ to the prompt background becomes more pronounced with \Wpm\Wpm\ being the largest contribution to the prompt background for $m_{\ell \ell} > 600$ \gev\ in the \mumu\ channel.

\begin{table*}[!ht]
\begin{center}
\begin{tabular}{l|c|c|c|c|c|r}
\hline
&\multicolumn{6}{c}{Number of electron pairs} \\
{$m(e^{\pm}e^{\pm})$} & \multirow{2}{*}{Prompt} & Non- &$e^{\pm}$ charge &$W\gamma \rightarrow$ & \multirow{2}{*}{Total SM}& \multirow{2}{*}{Data}\\
$[$\gev$]$& & Prompt & misid.&$Wee$ & & \\
\hline
 $>15$ & $ 347 \pm 25 $   &$ 520 \pm 120 $  &$ 1020 \pm 150 $ &$180 \pm 40 $   &$ 2060 \pm 190 $ &$ 1976 $\\
$>100$ &$ 174 \pm 14 $    &$ 250 \pm 50 $	 &$ 550 \pm 80 $   &$75 \pm 16 $&$ 1050 \pm 100 $  &$ 987 $\\
$>200$ &$ 51.5 \pm 4.9 $  &$ 72 \pm 13 $&$ 150 \pm 27 $   &$22 \pm 5 $ &$ 296 \pm 31 $   &$ 265 $\\
$>300$ & $ 15.7 \pm 1.9 $ &$ 23 \pm 5 $ &$ 43 \pm 12 $&$8.0 \pm 2.3 $  &$ 89 \pm 14 $&$ 83 $\\
$>400$ &$ 5.3 \pm 0.9 $   &$ 8.1 \pm 2.4 $  &$ 16 \pm 8 $ &$ 3.8 \pm 1.3 $ &$ 33 \pm 8 $ &$ 30 $\\
$>500$ &$ 2.3 \pm 0.5 $   &$ 3.1 \pm 1.5 $  &$ 6 \pm 5 $  &$ 2.7 \pm 1.0 $ &$ 14 \pm 5 $ &$ 13 $\\
$>600$ &$ 0.91 \pm 0.28 $   &$ 0.8 ^{+1.0}_{-0.8} $  &$ 6 \pm 5 $  &$ 1.0 \pm 0.6 $ &$ 9 \pm 5 $  &$ 7 $\\
\hline\hline
&\multicolumn{6}{c}{Number of electron--muon pairs} \\
{$m(e^{\pm}\mu^{\pm})$} & \multirow{2}{*}{Prompt} & Non- &$e^{\pm}$ charge &$W\gamma \rightarrow$ & \multirow{2}{*}{Total SM}& \multirow{2}{*}{Data}\\
$[$\gev$]$& & Prompt & misid.&$Wee$ & & \\
\hline
$>15$  &$1030 \pm 50$  &$910 \pm 220$ &$370 \pm 40$  &$270 \pm 50 $& $2580 \pm 240$ & $2315$\\
$>100$ &$458 \pm 26$   &$340 \pm 80$  &$87 \pm 11$   &$104 \pm 20 $& $990 \pm 90$&$859 $\\
$>200$ &$130 \pm 9$&$79 \pm 17$   &$29 \pm 4$&$ 28 \pm 6$& $265 \pm 22$  &$226 $ \\
$>300$ &$43 \pm 5$ &$24 \pm 6$&$9.5 \pm 1.9$ &$8.1 \pm 2.4  $& $84 \pm 8$&$85 $\\
$>400$ &$16.0 \pm 2.1$ &$9.2 \pm 3.0$ &$2.5 \pm 0.8$ &$2.7 \pm 1.1$& $31 \pm 4$  &$31 $\\[+0.025in]
$>500$ &$6.8 \pm 1.1$  &$2.8 \pm 1.5$ &$1.5 \pm 0.4$ &$ 1.6 \pm 0.8$& $12.6 \pm 2.1$ &$13 $\\[+0.025in]
$>600$ &$3.5 \pm 0.7$  &$1.6 \pm 1.0$ &$0.9 \pm 0.4$ &$ 1.2 \pm 0.7 $& $7.4 \pm 1.5$ &$9 $\\[+0.025in]
\hline\hline
&\multicolumn{4}{c}{Number of muon pairs} \\
{$m(\mu^{\pm}\mu^{\pm})$} & \multirow{2}{*}{Prompt} & Non- & \multirow{2}{*}{Total SM}& \multicolumn{1}{c}{\multirow{2}{*}{Data}} &\multicolumn{2}{c}{\multirow{5}{*}{}} \\
$[$\gev$]$& & Prompt & & \multicolumn{1}{c}{}&\multicolumn{2}{c}{}\\
\cline{1-5}
$>15$  &$580 \pm 40$  &$203 \pm 34$     &$780 \pm 50$   &  \multicolumn{1}{c}{$843 $}&\multicolumn{2}{c}{}\\ 
$>100$ &$245 \pm 21$  &$56 \pm 11$      &$301 \pm 24$   & \multicolumn{1}{c}{$330 $}&\multicolumn{2}{c}{}\\
$>200$ &$67 \pm 7$&$8.7 \pm 2.3$    &$76 \pm 8$ & \multicolumn{1}{c}{$87 $}&\multicolumn{2}{c}{}\\
$>300$ &$20.7 \pm 2.9$&$1.9 \pm 1.0$    & $22.6 \pm 3.1$& \multicolumn{1}{c}{$27 $}&\multicolumn{2}{c}{}\\
$>400$ &$7.7 \pm 1.5$ &$1.2 \pm 0.9$ & $9.0 \pm 1.7$ & \multicolumn{1}{c}{$9 $}&\multicolumn{2}{c}{}\\
$>500$ &$2.9 \pm 0.8$ &$0.32^{+0.41}_{-0.32}$ &$3.2 \pm 0.9$  & \multicolumn{1}{c}{$4 $}&\multicolumn{2}{c}{}\\
$>600$ &$0.9 \pm 0.4$ &$0.0^{+0.2}_{-0.0}$ &$0.9 \pm 0.4$  & \multicolumn{1}{c}{$1 $}&\multicolumn{2}{c}{}\\
\cline{1-5}
\end{tabular}
\end{center}
\caption{Expected and observed numbers of isolated same-sign lepton pairs in the \ee, \emu\ and \mumu\ channel for various cuts on the dilepton invariant mass, $m(\ell^{\pm} \ell^{\pm})$. The uncertainties shown are the systematic uncertainties.}
\label{tab:SMbackground}
\end{table*}

Table \ref{tab:SMbackground2} shows a similar comparison of the data with the SM expectation separately for $\ell^{+} \ell^+$ and $\ell^{-} \ell^-$ pairs. 
Due to the contribution of the valence quarks in the proton, more $W^+$ than $W^-$  bosons are produced in $pp$ collisions resulting in 
a higher background for the $\ell^+ \ell^+$ final state. 
For all final states no significant excesses or deficits are observed between the data and the SM background predictions.

\begin{table}[htpb]
\begin{center}
\begin{tabular}{l|cr|cr|cr}
\hline
{$m(\ell \ell)$} & \multicolumn{2}{c}{$e^+ e^+$ pairs}&\multicolumn{2}{c}{$e^+ \mu^+$ pairs}& \multicolumn{2}{c}{$\mu^+ \mu^+$ pairs}\\
$[$\gev$]$& Total SM & Data & Total SM & Data & Total SM & Data \\
\hline 
$>15$ &$ 1120 \pm 100 $ &$ 1124$&$1440 \pm 130$&$1327$&$454 \pm 32$   & 502\\
$>100$ &$ 610 \pm 60 $  &$ 593 $&$570 \pm 50$  &$523$ &$184 \pm 16$   & 198\\
$>200$ &$ 187 \pm 22 $ &$ 167 $ &$146 \pm 13$  &$143$ &$48 \pm 6$ & 62\\
$>300$ &$ 61 \pm 11$&$ 48 $  &$50 \pm 5$&$56$  & $15.3 \pm 2.2$&18\\
$>400$ &$ 19 \pm 6 $&$ 18 $&$18.4 \pm 2.6$&$21$  & $6.2 \pm 1.2$ &6\\
$>500$ &$ 9 \pm 5 $ &$ 9 $  &$7.8 \pm 1.4$ & $8 $ &$2.6 \pm 0.8$  &1\\
$>600$ & $ 7 \pm 5 $&$ 5 $   &$4.8 \pm 1.1$ &$6$   & $0.8 \pm 0.4$ &0\\[+0.025in]
\hline\hline
{$m(\ell \ell)$} & \multicolumn{2}{c}{$e^- e^-$ pairs}&\multicolumn{2}{c}{$e^- \mu^-$ pairs}& \multicolumn{2}{c}{$\mu^- \mu^-$ pairs}\\
$[$\gev$]$& Total SM & Data & Total SM & Data & Total SM & Data \\
\hline 
$>15$ &$ 940 \pm 100 $ &$ 852 $&$1140 \pm 110$   &$988 $ & $328 \pm 23$       & 341\\
$>100$ &$ 440 \pm 50 $  &$ 394 $&$417 \pm 40$     & $336 $&$117 \pm 9$         &132\\
$>200$ &$ 109 \pm 16 $  &$ 98 $  &$119 \pm 11$     & $83$  & $27.6 \pm 2.8$     &25\\
$>300$ &$ 29 \pm 7 $&$ 35 $&$35 \pm 4$   &$29$   & $7.3 \pm 1.2$      &9\\
$>400$ &$ 14 \pm 5 $&$ 12 $  &$12.1 \pm 2.3$   &$10 $  &$2.7 \pm 0.7$       &3\\
$>500$ &$ 5.0\pm 1.3 $  &$ 4 $  &$4.9 \pm 1.5$    &$5 $   &$0.64^{+0.33}_{-0.26}$       &3\\
$>600$ &$ 2.7 \pm 0.9 $ &$ 2 $ & $2.5 \pm 1.0$   &$3$    &$0.09^{+0.23}_{-0.09}$ &1\\[+0.025in]
\hline
\end{tabular}
\end{center}
\caption{Expected and observed numbers of positively or negatively charged lepton pairs for various cuts on the dilepton invariant mass, $m(\ell \ell)$. The uncertainties shown are the systematic uncertainties.}
\label{tab:SMbackground2}
\end{table}

Based on the above findings, upper limits are computed at the 95\% confidence level (CL) using the CL$_{\mathrm{S}}$~\cite{cls} prescription. Limits are given on the number of same-sign lepton pairs ($N_{95}$) contributed by new physics beyond the SM for various invariant mass thresholds. 
In this procedure the number of pairs in each mass bin is described using a Poisson probability density function. The systematic uncertainties (as discussed in section \ref{sec:uncertainties}) are incorporated into the limit calculation as nuisance parameters with Gaussian priors with the correlations between uncertainties taken into account. The limits can be translated into an upper limit on the 
fiducial cross-section using: $\sigma_{95}^{\mathrm{fid}} = N_{95}/(\epsilon_{\mathrm{fid}} \times \int L \mathrm{d}t)$, where $\epsilon_{\mathrm{fid}}$ is the efficiency for finding a lepton pair from a possible signal from new physics in the fiducial region at particle level, and $\int L \mathrm{d}t$ is the integrated luminosity. The efficiency $\epsilon_{\mathrm{fid}}$ is discussed in detail in the next section.
\begin{table}[ht]
  \begin{center}
    \begin{tabular}{l|c|c}
\hline
 Selection     & Electron requirement & Muon requirement\\
\hline
Leading lepton \pt & $\pt > 25$ \gev & $\pt > 25$ \gev\\
Subleading lepton \pt & $\pt > 20$ \gev & $\pt > 20$ \gev\\
Lepton $\eta$ & $|\eta| < 1.37$ or $1.52 < |\eta| < 2.47$ & $|\eta| < 2.5$\\
Isolation & $\sum \pt (\Delta R=0.3)/\pt^{\mathrm{e}} < 0.1$& $\sum \pt (\Delta R=0.3)/\pt^{\mu} < 0.07$\\
      \hline\hline
 Selection     & \multicolumn{2}{c}{Event selection}\\
\hline
Lepton pair & \multicolumn{2}{c}{Same-sign pair with $m_{\ell \ell} > 15$~\gev}\\
Electron pair & \multicolumn{2}{c}{Veto pairs with $70 < m_{\ell \ell} < 110$~\gev}\\
Event & \multicolumn{2}{c}{No opposite-sign same-flavour pair with $|m_{\ell \ell} - m_{Z}| < 10$~\gev}\\  
\hline

\end{tabular}
\end{center}
  \caption{Summary of requirements on generated leptons and lepton pairs in the fiducial region at particle level. More information on the calculation of the isolation \pt\ is given in the text.
  }
\label{tab:fidcuts}
\end{table}

The fiducial volume at particle level, as summarised in table \ref{tab:fidcuts}, is chosen to be very similar to the one used in the object and event selections (see sections \ref{sec:particles} and \ref{sec:selection}). The leptons must be isolated and fulfil the same kinematic requirements on transverse momentum and pseudorapidity as imposed at reconstruction level. 
Lepton isolation is implemented by requiring that the sum
of the \pt\ of the stable charged particles with $\pt> 1~(0.4)$~\gev\ in a cone of size $\Delta R=0.3$ around the lepton is required to be less than 7\% (10\%) of the lepton \pt\ for muons (electrons).
In addition, the two leptons must have the same charge and pass the same invariant mass cut, $m_{\ell \ell} > 15$~\gev, as required at reconstruction level. 
In addition, in the \ee\ channel the mass range $70 < m_{\ell \ell} < 110$~\gev\ is vetoed. 
Finally,
events are rejected in which an opposite-sign, same-flavour lepton pair is found with $|m_{\ell \ell} - m_{Z}| < 10$~\gev.

\subsection{Fiducial cross-section limits}
\label{sec:incllimits}
To derive upper limits on the cross-section due to physics beyond the SM, the fiducial efficiency, $\epsilon_{\mathrm{fid}}$, is calculated. The quantity $\epsilon_{\mathrm{fid}}$ is the ratio, for leptons from the signal
processes, of the number of selected lepton pairs to the number of
true same-sign lepton pairs satisfying the fiducial selection at particle
level.
The value of $\epsilon_{\mathrm{fid}}$ generally depends on the new physics process, e.g.~the number of leptons in the final state passing the kinematic selection criteria or the number of jets that may affect the lepton isolation. To minimise this dependence, the definition of the fiducial region is closely related to the analysis selection.
The limits are quoted using the lowest fiducial efficiency obtained for the following beyond-the-SM processes. 

Firstly, production of doubly charged Higgs boson pairs with masses ranging between 100 \gev\ and 1 \tev\ is considered.
Another process is production of a diquark ($S_{\mathrm{DQ}}$) with charge $\pm 2/3$ or $\pm 4/3$ in the Zee--Babu model, which decays into two same-sign leptoquarks ($S_{\mathrm{LQ}}$), decaying subsequently into a same-sign lepton pair. This model is considered for diquark masses between 2.5 \tev\ and 3.5 \tev\ and leptoquark masses between 1 \tev\ and 1.4 \tev.
The third process is a production of a right-handed $W_{\mathrm{R}}$ boson decaying into a lepton and a Majorana neutrino $N_{\mathrm{R}}$, with $N_{\mathrm{R}}$ subsequently decaying into a lepton and two jets, for $W_{\mathrm{R}}$ masses between 1 \tev\ and 2 \tev, and $N_{\mathrm{R}}$ masses between 250 \gev\ and 1.5 \tev. 
The last process is pair production of $b'$ chiral quarks, decaying either exclusively into $Wt$ or decaying into $Wq$, $q$ being an up-type quark, with a 33\% branching ratio in each quark channel, for $b'$ masses between 400~\gev\ and 1~\tev. 

The fiducial efficiencies vary between 46\% and 74\% with similar values 
for the \ee, \emu\ and \mumu\ final states.
The lowest values of $\epsilon_{\mathrm{fid}}$ are found in the case of the fourth-generation down-type chiral quark model, and the highest for the production of $W_{\mathrm{R}}$ bosons and $N_{\mathrm{R}}$ neutrinos. 
The primary reason for this dependence is that the electron identification efficiency varies by about 15\% over the relevant \pt\ range \cite{elepaper}. For muons differences in the efficiencies arise due to the detector acceptances \cite{mupaper}, which are populated differently depending on the kinematics of the leptons produced in the new physics process.
Further differences of the order of 1\% arise since the calorimeter-based isolation criterion is not emulated because the isolation energy has a poor resolution in the calorimeter.
The fiducial efficiencies are also derived separately for $\ell^+ \ell^+$ and $\ell^- \ell^-$ pairs and found to be charge independent.

\begin{figure}[!ht]
\begin{center}
\subfigure[]{
  \includegraphics[width=0.58\columnwidth]{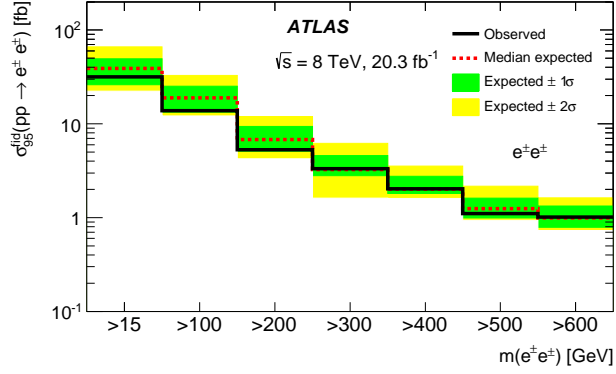}
}
\subfigure[]{
  \includegraphics[width=0.58\columnwidth]{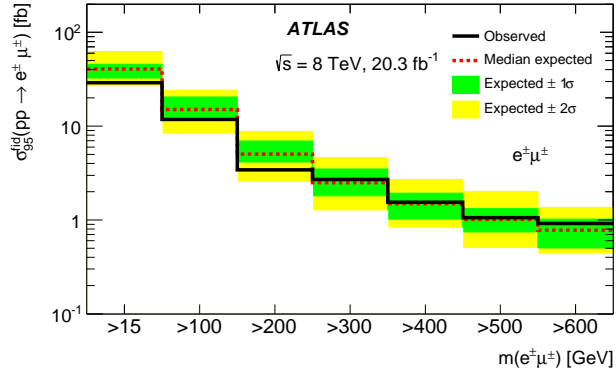}
}
\subfigure[]{
  \includegraphics[width=0.58\columnwidth]{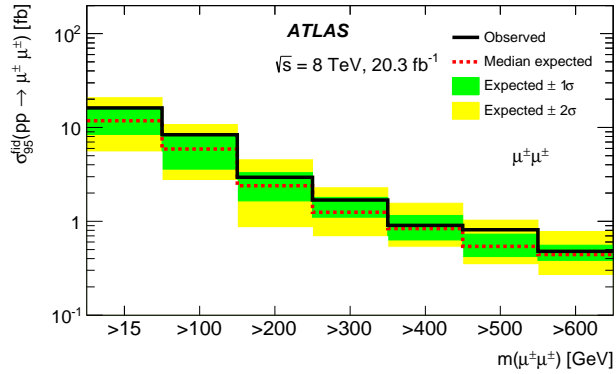}
}
\caption{\label{fig:incllimit} 
Fiducial cross-section limits at 95\% CL
for (a) \ee, (b) \emu\ and (c) \mm\ pairs, as a function of the lower bound on the lepton pair mass. 
The green and yellow bands show the $1\sigma$ and $2\sigma$ bands on the expected limits. The mass range between 70~\gev\ and 110~\gev\ is not included in the \ee\ channel as this region is used to estimate the background from charge misidentification.
 }
\end{center}
\end{figure}
In the following, cross-section limits are presented for the case which yields the lowest fiducial efficiency, that is 48\%, 50\% and 46\% in the \ee, \emu\ and \mumu\ channel respectively. The 95\% CL upper limits on the fiducial cross-section are shown in figure \ref{fig:incllimit} and in table \ref{tab:incllimit} separately for each final state.
\begin{table*}[!ht]
\begin{center}
\begin{tabular}{c||c|c||c|c||c|c}
\hline
 & \multicolumn{6}{c}{95\%  CL upper limit [fb]} \\
 & \multicolumn{2}{c||}{$e^{\pm}e^{\pm}$} & \multicolumn{2}{c||}{$e^{\pm}\mu^{\pm}$} & \multicolumn{2}{c}{$\mu^{\pm}\mu^{\pm}$} \\
Mass range & Expected & Observed & Expected & Observed & Expected & Observed \\
\hline
\rule{0pt}{3ex}
$>15$~GeV & $39^{+10}_{-13}$ & 32 & $41^{+5}_{-8}$ & 29 & $12^{+4}_{-3}$ & 16
 \\[+0.05in]
$>100$~GeV & $19^{+6}_{-6}$ & 14 & $15.1^{+5.5}_{-2.6}$ & 11.8 & $5.9^{+2.2}_{-2.3}$ & 8.4
 \\[+0.05in]
$>200$~GeV & $6.8^{+2.6}_{-1.7}$ & 5.3 & $5.0^{+1.9}_{-0.9}$ & 3.4 & $2.4^{+0.9}_{-0.8}$ & 2.9
 \\[+0.05in]
$>300$~GeV & $3.3^{+1.3}_{-0.4}$ & 3.3 & $2.5^{+1.0}_{-0.7}$ & 2.7 & $1.25^{+0.55}_{-0.15}$ & 1.69
 \\[+0.05in]
$>400$~GeV & $2.02^{+0.74}_{-0.21}$ & 2.03 & $1.5^{+0.4}_{-0.5}$ & 1.6 & $0.83^{+0.32}_{-0.20}$ & 0.91
 \\[+0.05in]
$>500$~GeV & $1.25^{+0.36}_{-0.26}$ & 1.10 & $1.02^{+0.30}_{-0.27}$ & 1.06 & $0.54^{+0.19}_{-0.12}$ & 0.82
 \\[+0.05in]
$>600$~GeV & $0.99^{+0.34}_{-0.20}$ & 1.02 & $0.78^{+0.24}_{-0.28}$ & 0.92 & $0.44^{+0.11}_{-0.06}$ & 0.48
 \\[+0.05in]
\hline\hline
 & \multicolumn{2}{c||}{$e^{+}e^{+}$} & \multicolumn{2}{c||}{$e^{+}\mu^{+}$} & \multicolumn{2}{c}{$\mu^{+}\mu^{+}$} \\
Mass range & Expected & Observed & Expected & Observed & Expected & Observed \\
[+0.05in]
\hline
\rule{0pt}{3ex}
$>15$~GeV & $27^{+11}_{-6}$ & 28 & $25^{+10}_{-4}$ & 23 & $9.5^{+3.3}_{-3.1}$ & 14
 \\[+0.05in]
$>100$~GeV & $14.3^{+5.4}_{-2.8}$ & 13.5 & $11^{+4}_{-2.1}$ & 9 & $5.0^{+1.6}_{-1.3}$ & 6.3
 \\[+0.05in]
$>200$~GeV & $5.4^{+2.0}_{-1.4}$ & 4.6 & $3.6^{+1.3}_{-0.7}$ & 3.6 & $2.2^{+0.8}_{-0.5}$ & 3.6
 \\[+0.05in]
$>300$~GeV & $2.5^{+0.9}_{-0.6}$ & 2.0 & $1.9^{+0.8}_{-0.5}$ & 2.6 & $1.11^{+0.46}_{-0.29}$ & 1.42
 \\[+0.05in]
$>400$~GeV & $1.59^{+0.47}_{-0.34}$ & 1.64 & $1.10^{+0.46}_{-0.23}$ & 1.39 & $0.74^{+0.27}_{-0.17}$ & 0.74
 \\[+0.05in]
$>500$~GeV & $1.44^{+0.34}_{-0.36}$ & 1.55 & $0.79^{+0.21}_{-0.22}$ & 0.89 & $0.42^{+0.24}_{-0.10}$ & 0.38
 \\[+0.05in]
$>600$~GeV & $1.27^{+0.37}_{-0.26}$ & 1.10 & $0.65^{+0.14}_{-0.16}$ & 0.77 & $0.37^{+0.09}_{-0.05}$ & 0.32
 \\[+0.05in]
\hline\hline
 & \multicolumn{2}{c||}{$e^{-}e^{-}$} & \multicolumn{2}{c||}{$e^{-}\mu^{-}$} & \multicolumn{2}{c}{$\mu^{-}\mu^{-}$} \\
Mass range & Expected & Observed & Expected & Observed & Expected & Observed \\
[+0.05in]
\hline
\rule{0pt}{3ex}
$>15$~GeV & $23^{+8}_{-5}$ & 19 & $19.0^{+8.0}_{-2.8}$ & 16.0 & $6.8^{+2.7}_{-1.5}$ & 8.3
 \\[+0.05in]
$>100$~GeV & $10.8^{+4.4}_{-2.4}$ & 9.0 & $8.2^{+2.2}_{-2.1}$ & 5.6 & $3.5^{+1.4}_{-0.9}$ & 5.1
 \\[+0.05in]
$>200$~GeV & $3.9^{+1.4}_{-1.2}$ & 3.5 & $2.8^{+1.2}_{-0.9}$ & 1.5 & $1.41^{+0.54}_{-0.33}$ & 1.29
 \\[+0.05in]
$>300$~GeV & $2.1^{+0.7}_{-0.5}$ & 2.6 & $1.6^{+0.6}_{-0.4}$ & 1.3 & $0.79^{+0.30}_{-0.16}$ & 1.0
 \\[+0.05in]
$>400$~GeV & $1.56^{+0.41}_{-0.31}$ & 1.35 & $0.91^{+0.34}_{-0.26}$ & 0.77 & $0.52^{+0.20}_{-0.13}$ & 0.59
 \\[+0.05in]
$>500$~GeV & $0.69^{+0.27}_{-0.17}$ & 0.64 & $0.62^{+0.12}_{-0.12}$ & 0.65 & $0.355^{+0.139}_{-0.013}$ & 0.683
 \\[+0.05in]
$>600$~GeV & $0.58^{+0.21}_{-0.08}$ & 0.61 & $0.49^{+0.16}_{-0.10}$ & 0.59 & $0.332^{+0.014}_{-0.011}$ & 0.454
 \\[+0.05in]
\hline

\end{tabular}
\end{center}
 \caption{Upper limit at 95\% CL on the fiducial cross-section for $\ell^{\pm} \ell^{\pm}$ pairs from non-SM signals. The expected limits and their $1 \sigma$ uncertainties are given together with the observed limits derived from the data. Limits are given separately for the \ee, \emu\ and \mumu\ channel inclusively and separated by charge.}
\label{tab:incllimit}
\end{table*}
The cross-section limits are statistical combinations of the $\ell^+ \ell^+$ and $\ell^- \ell^-$ limits and observed limits vary between 0.48~fb and 32~fb depending on the mass cut and the final state for the inclusive analysis. 
The limits obtained for $\ell^+ \ell^+$ and $\ell^- \ell^-$ pairs are also shown in table \ref{tab:incllimit} and range between 0.32~fb to 28~fb. 
Since the total limits are the limits on the sum of $\ell^+ \ell^+$ and $\ell^- \ell^-$, they are, in general, larger than charge separated limits. 
For all final states the observed limits are generally within $1 \sigma$ of the expected limits, which are obtained using simulated pseudo-experiments using only SM processes.

\subsection{Cross-section and mass limits for pair-produced doubly charged Higgs bosons}
\label{sec:dch}
As an example of the models producing same-sign lepton pairs, mass limits are obtained for doubly charged Higgs bosons, 
 which are pair produced via $s$-channel $Z$ boson or photon exchange in the framework of the left-right symmetric models \cite{LRSM1, LRSM2, LRSM3, LRSM4}.  
In this framework, left-handed states, $H_{\mathrm{L}}^{\pm \pm}$, and right-handed states $H_{\mathrm{R}}^{\pm \pm}$ are predicted. These Higgs bosons have identical kinematic properties, but their production rate differs due to the different couplings to $Z$ bosons \cite{Muhlleitner:2003me}. The cross-section, which is known at NLO, is around 2.5 times higher for $H_{\mathrm{L}}^{++} H_{\mathrm{L}}^{--}$ pair production compared to $H_{\mathrm{R}}^{++} H_{\mathrm{R}}^{--}$. 
In this analysis the decay of the two $H^{\pm \pm}$ bosons into leptons ($H^{\pm \pm} H^{\mp \mp} \rightarrow \ell_1^{\pm} \ell_2^{\pm} \ell_3^{\mp} \ell_4^{\mp}$) is considered. This is done using the same search strategy as for the fiducial cross-section limits, which looks for signs of new physics in events containing same-sign lepton pairs. 
Alternatively this search could be carried out looking for events with two same-sign lepton pairs (four-lepton final states). 
However, the four-lepton channel has a low efficiency 
due to the cases where at least one of the leptons falls outside the acceptance.

In the following, $H^{\pm\pm}$ boson mass values in the range 50 \gev\ to 1 \tev\ are considered. 
The cross-section is determined using $\sigma_{HH} \times BR = N_{H}^{\mathrm{rec}}/(2 \times A \times \epsilon \times \int L \mathrm{d}t)$, where $BR$ is the branching ratio of the decay into a lepton pair ($H^{\pm \pm} \rightarrow \ell^{\pm} \ell'^{\pm}$), $N_{H}^{\mathrm{rec}}$ is the number of reconstructed Higgs boson candidates, $A \times \epsilon$ is the acceptance times efficiency to find a lepton pair from the $H^{\pm \pm}$ decay, and the factor of two accounts for the two same-sign lepton pairs from the $H^{++}$ and $H^{--}$ bosons. The $A \times \epsilon$ is calculated for the simulated mass points and masses in between are interpolated via an empirical fit function. 
In the mass range considered in this analysis, the width of the $H^{\pm \pm}$ resonance is much smaller than the detector resolution of the lepton pairs. To extract the cross-section limits of $H^{\pm \pm}$ bosons the size of the mass bins used is optimised for each final state, such that 
in each mass bin the Higgs selection efficiency is very similar. 
Limits on the cross-section for pair production of
$H^{\pm \pm}$ and $H^{\mp \mp}$
bosons times the branching ratio in each of the three final states are extracted using the CL$_{\mathrm{S}}$ technique.
\begin{figure*}[!hp]
\begin{center}
\subfigure[]{
  \includegraphics[width=0.53\columnwidth]{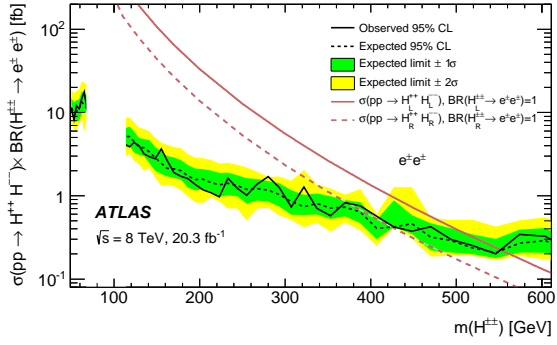}
}
\subfigure[]{
  \includegraphics[width=0.53\columnwidth]{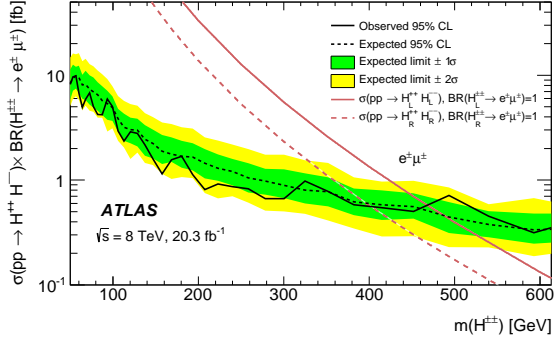}
}
\subfigure[]{
  \includegraphics[width=0.53\columnwidth]{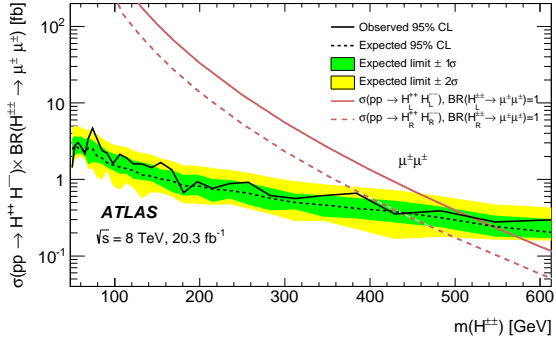}
}
\caption{\label{fig:dch} Upper limits at 95\% CL on the cross-section as a function of the dilepton invariant mass for the production of a doubly charged Higgs boson decaying into (a) \ee, (b) \emu, and (c) \mm\ pairs with a branching ratio of 100\%. The green and yellow bands correspond to the $1\sigma$ and $2\sigma$ bands on the expected limits respectively. Also shown are the expected cross-sections as a function of mass for left- and right-handed \dch. The mass range between 70~\gev\ and 110~\gev\ is not included in the \ee\ channel as this region is used to estimate the background from charge misidentification.}
\end{center}
\end{figure*}
\afterpage{\clearpage}

The results at 95\% CL are shown in figure \ref{fig:dch}. The scatter between adjacent mass bins in the observed limits is due to fluctuations in the background yields derived from limited statistics.
In general, good agreement is seen between observed and expected limits with maximum deviations of $2\sigma$. 
In the three final states, the cross-section limits vary between 11 fb for a \dch\ mass of 50 \gev\ to around 0.3 fb for a \dch\ mass of 600 \gev.
The expected cross-section curves for the pair production of $H_{\mathrm{L}}$ and $H_{\mathrm{R}}$ are also shown in figure \ref{fig:dch}.
\begin{table}[!ht]
\begin{center}
\begin{tabular}{c||c|c||c|c||c|c}
\hline
 & \multicolumn{6}{c}{95\% CL lower limit [\gev]} \\
 & \multicolumn{2}{c||}{$e^{\pm}e^{\pm}$}
 & \multicolumn{2}{c||}{$e^{\pm}\mu^{\pm}$}
 & \multicolumn{2}{c}{$\mu^{\pm}\mu^{\pm}$} \\
Signal & Expected & Observed 
& Expected & Observed 
& Expected & Observed \\
[+0.04in]
\hline
\rule{0pt}{3ex}
$H_{\mathrm{L}}^{\pm \pm}$ & $553 \pm 30$ & $551$ & $487 \pm 41$ & $468$ & $543 \pm 40$ & $516$
 \\[+0.04in]
\hline
\rule{0pt}{3ex}
$H_{\mathrm{R}}^{\pm \pm}$ & $425 \pm 30$ & $374$ & $396 \pm 34$ & $402$ & $435 \pm 33$ & $438$
 \\[+0.04in]
\hline
\end{tabular}
\end{center}
\caption{Lower limits at 95\% CL on the mass of $H_{\mathrm{L}}^{\pm \pm}$ and $H_{\mathrm{R}}^{\pm \pm}$ bosons, assuming a 100\% branching ratio to \ee, \emu\ and \mumu\ pairs. The $1\sigma$ variations are also shown for the expected limits.}
\label{tab:dch}
\end{table}
The lower mass limits
are given by the crossing point of the cross-section limit curve and the expected curve, and are summarised in table~\ref{tab:dch}.
The $1\sigma$ errors on the expected limits are symmetrised to reduce the effect from bin by bin statistical fluctuations. 
For this scenario the best limits are obtained for $H_{\mathrm{L}}^{\pm \pm}$ in the \ee\ channel with a limit around 550 \gev\ and for $H_{\mathrm{R}}^{\pm \pm}$ in the \mumu\ channel with a limit around 430 \gev. The limits are 10--20\% worse in the \emu\ channel due to the larger background at high invariant masses from $WZ$ production. 
The $WZ$ gives approximately twice as many events in the \emu\ channel than in the \ee\ or \mumu\ channel whereas the signal contributions are similar in all three channels. 
The mass limit on the singlet $H^{\pm \pm}$ predicted in the Zee--Babu model~\cite{ZBM3} is the same as the one obtained for $H_{\mathrm{L}}^{\pm \pm}$ as the cross-sections and decay kinematics are identical. Compared to the results based on the 2011 data \cite{2011dch}, the limits on the doubly charged Higgs mass are increased by 30--40\%. The mass limits vary with the branching ratio of the \dch\ decay into lepton pairs. 
\begin{figure}[!ht]
\begin{center}
\subfigure[]{
  \includegraphics[width=0.479\columnwidth]{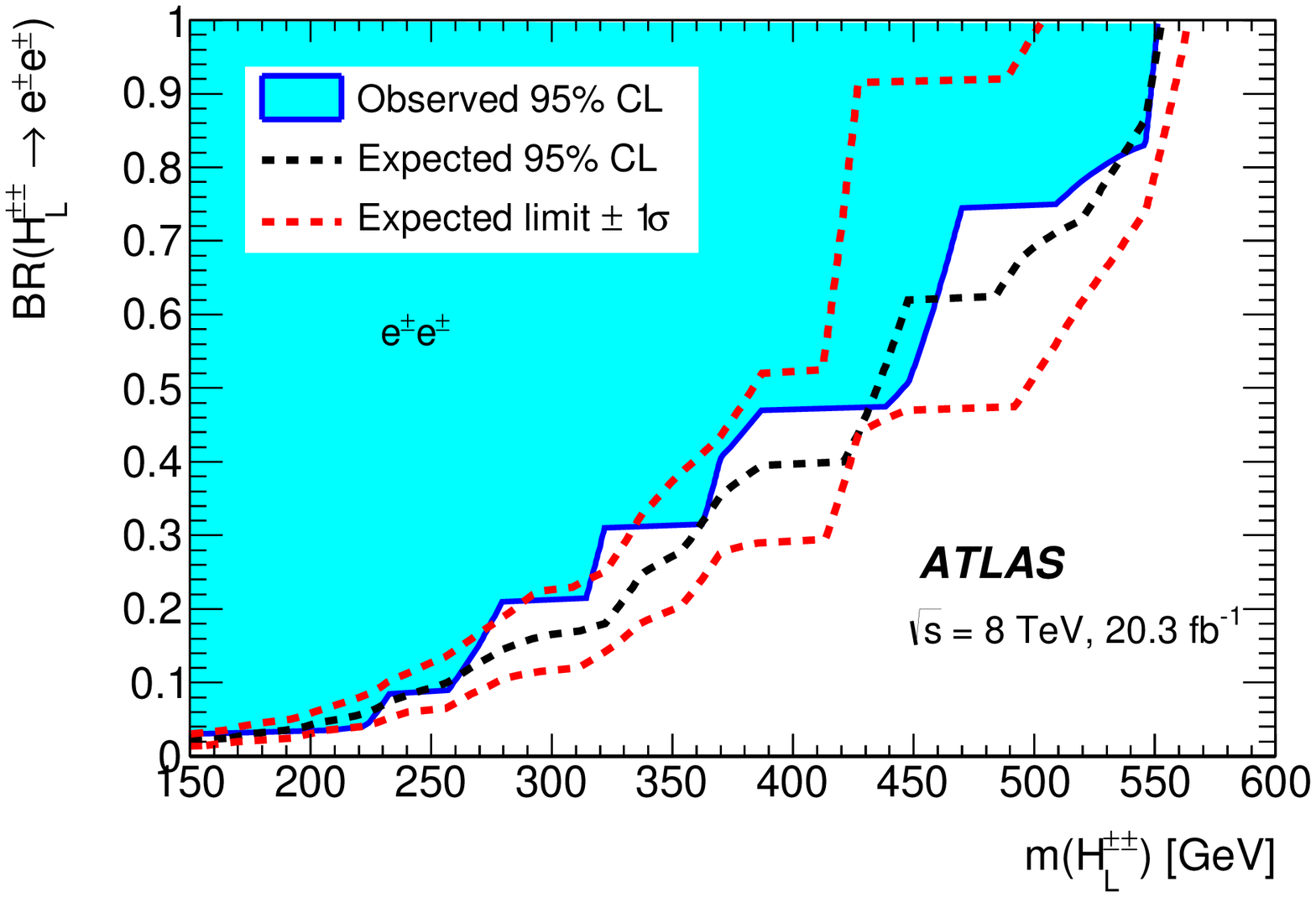}
}
\subfigure[]{
  \includegraphics[width=0.479\columnwidth]{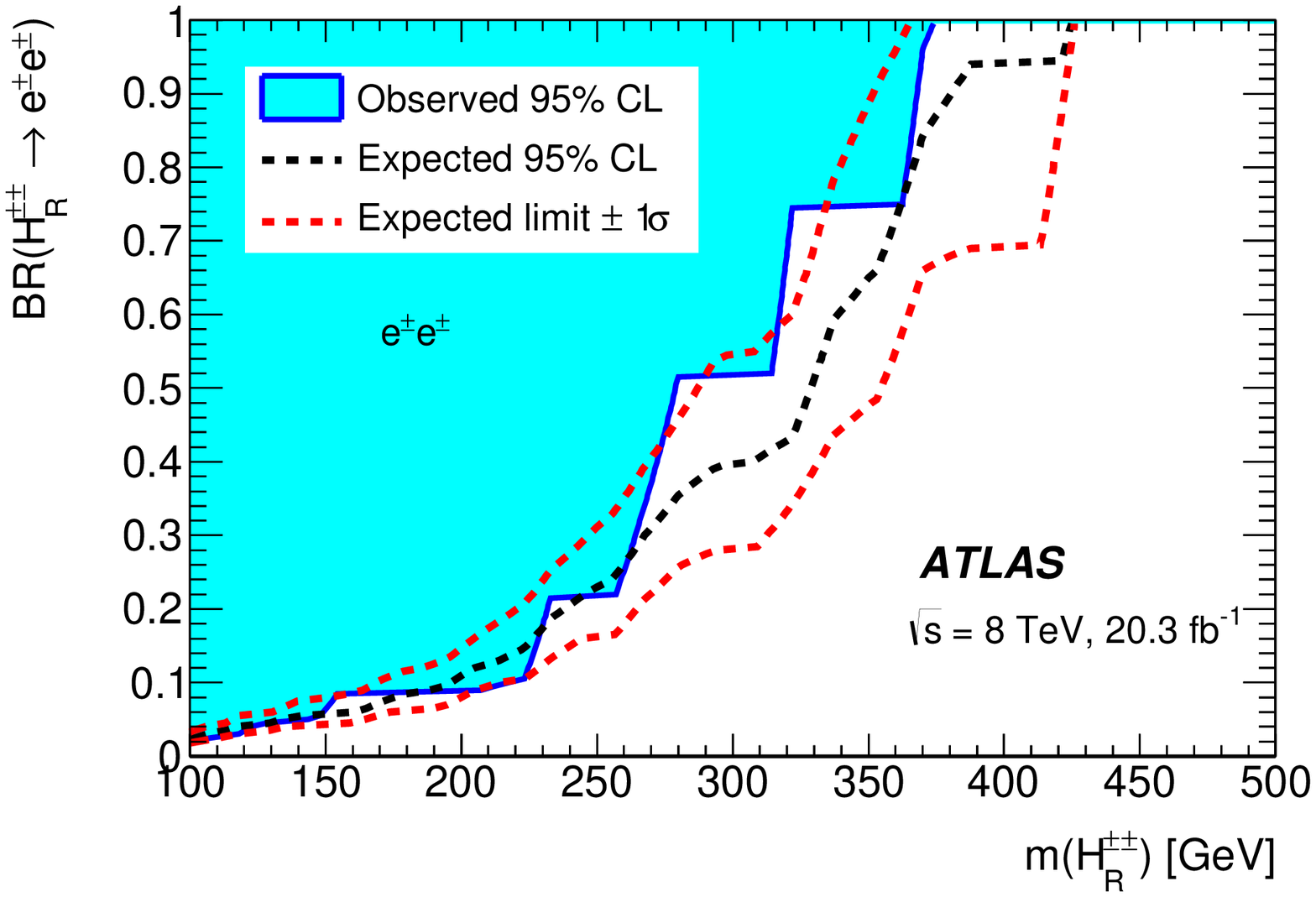}
}
\subfigure[]{
  \includegraphics[width=0.479\columnwidth]{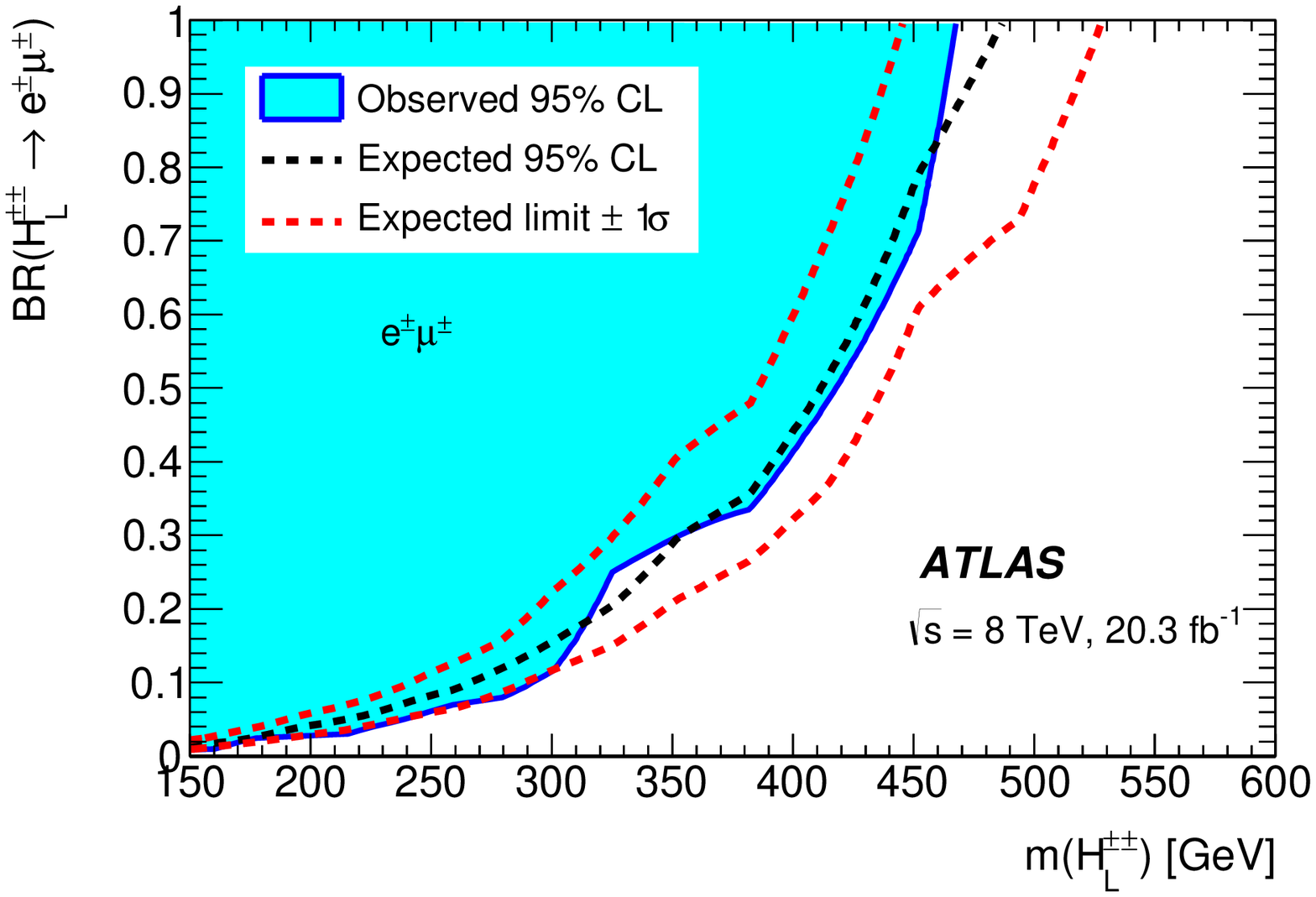}
}
\subfigure[]{
  \includegraphics[width=0.479\columnwidth]{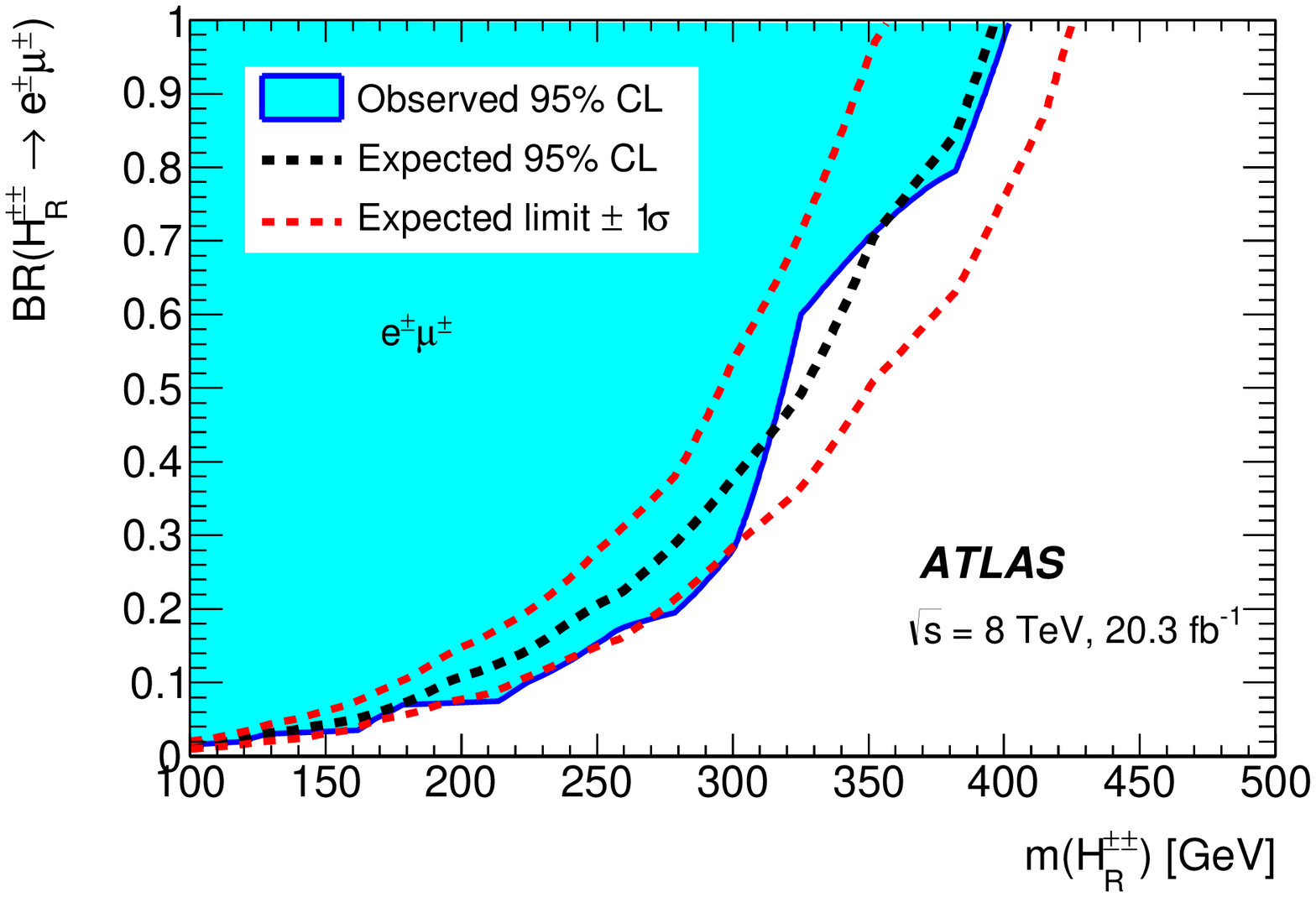}
}
\subfigure[]{
  \includegraphics[width=0.479\columnwidth]{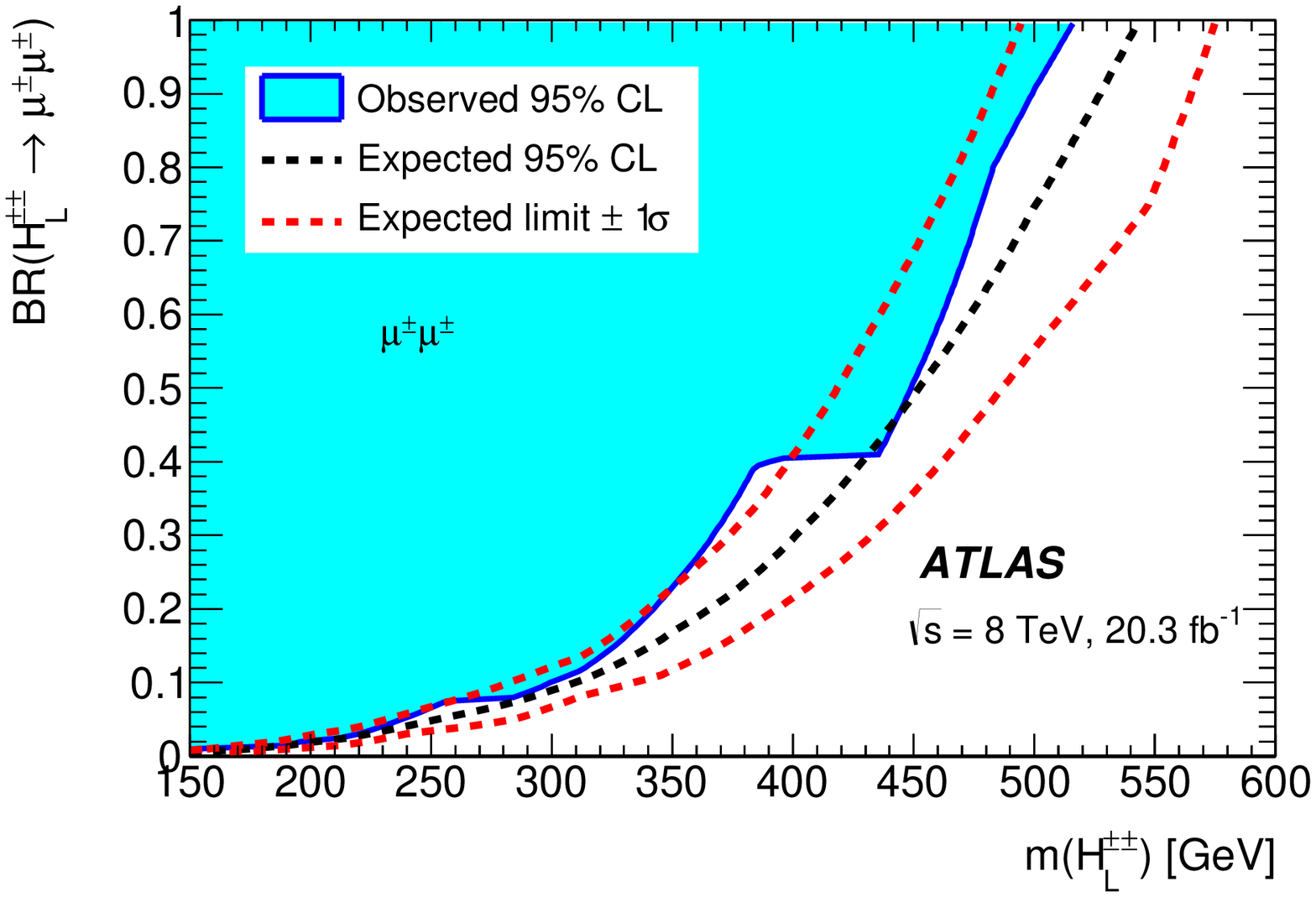}}
\subfigure[]{
  \includegraphics[width=0.479\columnwidth]{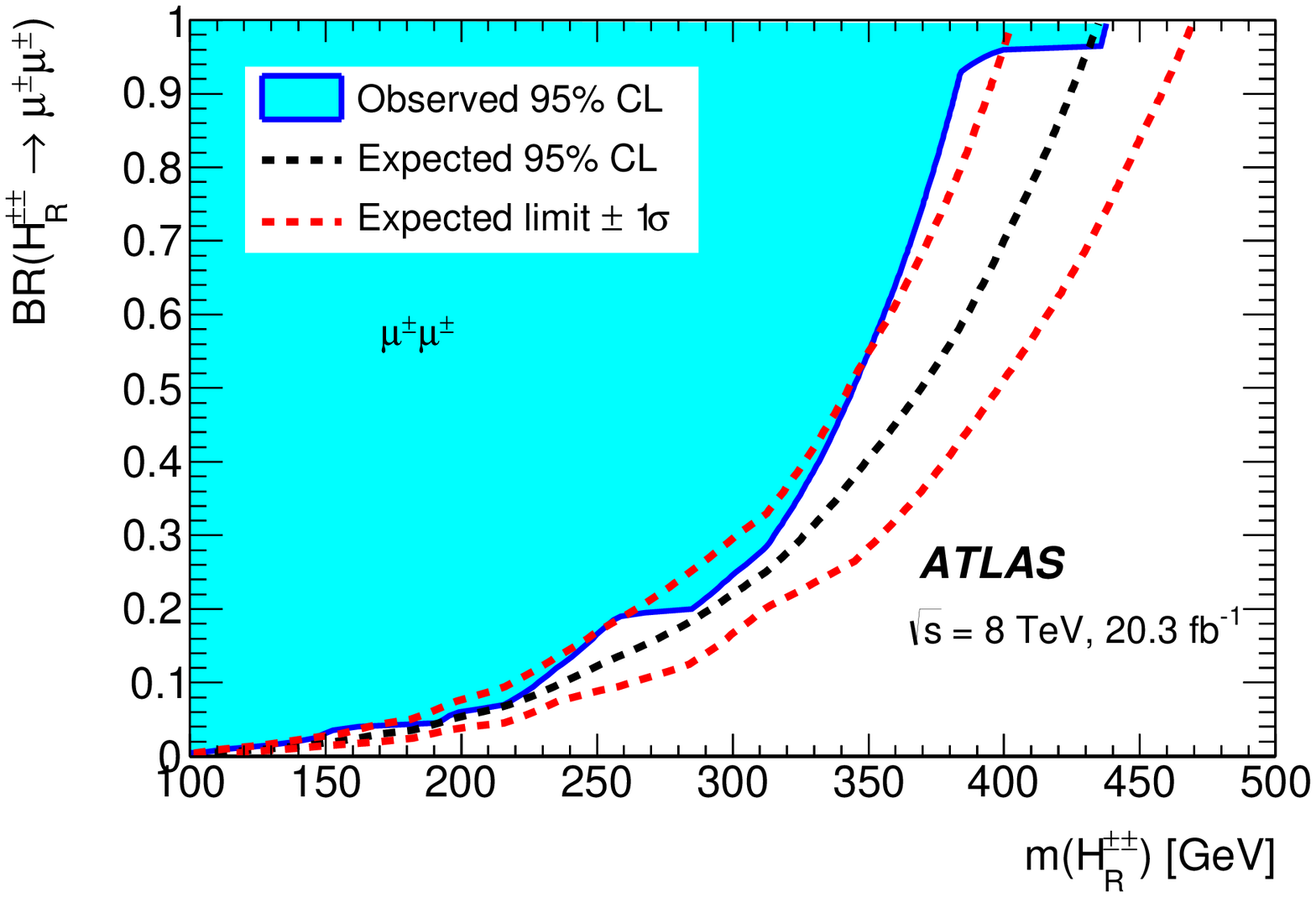}}
\caption{\label{fig:br}
Observed and expected 95\% CL limits on $H_{\mathrm{L}}^{\pm \pm} \rightarrow \ell^{\pm} \ell^{\pm}$ (left column) and $H_{\mathrm{R}}^{\pm \pm} \rightarrow \ell^{\pm} \ell^{\pm}$ production (right column) in the branching ratio versus $H^{\pm \pm}$ mass plane for the \ee\ (top), \emu\ (middle) and \mumu\ (bottom) channel. The blue shaded area is excluded.
}
\end{center}
\end{figure}
Figure \ref{fig:br} shows the mass limits as a function of the branching ratio for $H_{\mathrm{L}}^{\pm \pm}$ and $H_{\mathrm{R}}^{\pm \pm}$ in the three final states.

%% file: Conclusions.tex
In this paper a search for anomalous production of same-sign \ee, \emu, and \mumu\ pairs is presented using 20.3 \ifb\ of $\sqrt{s} =8$ \tev\ $pp$ collision data recorded with the ATLAS detector at the LHC. 
To make this search as inclusive as possible, there are no additional requirements on missing transverse momentum, jets, or other final-state particles.
The data agree with the SM expectation and no significant deviations are observed. Fiducial cross-section limits are derived for contributions from new physics beyond the SM, which give rise to final states with two same-sign isolated leptons.
The 95\% CL upper limits on the cross-section are given as functions of a threshold on the invariant mass of the lepton pair.
For invariant mass cuts ranging between 15 \gev\ and 600 \gev, the observed fiducial cross-section 
upper limit varies between 0.48 fb and 32 fb depending on the dilepton invariant mass and flavour combination. 
Limits are also set for a model of doubly charged Higgs bosons, in which doubly charged Higgs bosons are pair produced.
Assuming these Higgs bosons decay exclusively into \ee, \emu\ or \mumu\ pairs,
95\% CL lower mass limits for left-handed Higgs bosons between 465 \gev\ and 550 \gev\ are obtained depending on the flavour of the lepton pair. The mass limits for right-handed Higgs bosons range between 370 \gev\ and 435 \gev. These results represent a significant improvement compared to the previous ATLAS results based on $\sqrt{s} = 7$ \tev\ data.

%% file: Acknowledgement.tex
We thank CERN for the very successful operation of the
LHC, as well as the support staff from our institutions
without whom ATLAS could not be operated effciently.

We acknowledge the support of ANPCyT, Argentina;
YerPhI, Armenia; ARC, Australia; BMWF
and FWF, Austria; ANAS, Azerbaijan; SSTC, Belarus;
CNPq and FAPESP, Brazil; NSERC, NRC
and CFI, Canada; CERN; CONICYT, Chile; CAS,
MOST and NSFC, China; COLCIENCIAS, Colombia;
MSMT CR, MPO CR and VSC CR, Czech Republic;
DNRF, DNSRC and Lundbeck Foundation, Denmark;
EPLANET, ERC and NSRF, European Union;
IN2P3-CNRS, CEA-DSM/IRFU, France; GNSF, Georgia;
BMBF, DFG, HGF, MPG and AvH Foundation,
Germany; GSRT and NSRF, Greece; ISF, MINERVA,
GIF, I-CORE and Benoziyo Center, Israel;
INFN, Italy; MEXT and JSPS, Japan; CNRST, Morocco;
FOM and NWO, Netherlands; BRF and RCN,
Norway; MNiSW and NCN, Poland; GRICES and
FCT, Portugal; MNE/IFA, Romania; MES of Russia
and ROSATOM, Russian Federation; JINR; MSTD,
Serbia; MSSR, Slovakia; ARRS and MIZ\v{S}, Slovenia;
DST/NRF, South Africa; MINECO, Spain; SRC and
Wallenberg Foundation, Sweden; SER, SNSF and Cantons
of Bern and Geneva, Switzerland; NSC, Taiwan;
TAEK, Turkey; STFC, the Royal Society and Leverhulme
Trust, United Kingdom; DOE and NSF, United
States of America.

The crucial computing support from all WLCG
partners is acknowledged gratefully, in particular from
CERN and the ATLAS Tier-1 facilities at TRIUMF
(Canada), NDGF (Denmark, Norway, Sweden),
CC-IN2P3 (France), KIT/GridKA (Germany), INFNCNAF
(Italy), NL-T1 (Netherlands), PIC (Spain),
ASGC (Taiwan), RAL (UK) and BNL (USA) and in
the Tier-2 facilities worldwide.

%% file: atlas_authlist.tex
\begin{flushleft}
{\Large The ATLAS Collaboration}

\bigskip

G.~Aad$^{\rm 84}$,
B.~Abbott$^{\rm 112}$,
J.~Abdallah$^{\rm 152}$,
S.~Abdel~Khalek$^{\rm 116}$,
O.~Abdinov$^{\rm 11}$,
R.~Aben$^{\rm 106}$,
B.~Abi$^{\rm 113}$,
M.~Abolins$^{\rm 89}$,
O.S.~AbouZeid$^{\rm 159}$,
H.~Abramowicz$^{\rm 154}$
% ...
\bigskip
\\
% ...
$^{11}$ Institute of Physics, Azerbaijan Academy of Sciences, Baku, Azerbaijan\\
% ...
$^{84}$ CPPM, Aix-Marseille Universit{\'e} and CNRS/IN2P3, Marseille, France\\
% ...
$^{89}$ Department of Physics and Astronomy, Michigan State University, East Lansing MI, United States of America\\
% ...
$^{106}$ Nikhef National Institute for Subatomic Physics and University of Amsterdam, Amsterdam, Netherlands\\
% ...
$^{112}$ Homer L. Dodge Department of Physics and Astronomy, University of Oklahoma, Norman OK, United States of America\\
$^{113}$ Department of Physics, Oklahoma State University, Stillwater OK, United States of America\\
% ...
$^{116}$ LAL, Universit{\'e} Paris-Sud and CNRS/IN2P3, Orsay, France\\
% ...
$^{152}$ Institute of Physics, Academia Sinica, Taipei, Taiwan\\
% ...
$^{154}$ Raymond and Beverly Sackler School of Physics and Astronomy, Tel Aviv University, Tel Aviv, Israel\\
% ...
$^{159}$ Department of Physics, University of Toronto, Toronto ON, Canada\\
% ...
\end{flushleft}
